\pdfoutput=1
\documentclass[sigplan]{acmart}

\renewcommand\footnotetextcopyrightpermission[1]{}
\acmJournal{PACMPL}
\acmVolume{1}
\acmNumber{OOPSLA}
\acmArticle{1}
\acmYear{2019}
\acmMonth{1}
\acmDOI{} 
\startPage{1}

\setcopyright{none}

\usepackage{booktabs}   
\usepackage{subcaption} 

\usepackage{xspace}

\usepackage{pifont}
\usepackage{pgfplots}
\usepackage{pgf-pie}
\pgfplotsset{compat=1.12}
\usepgfplotslibrary{statistics}

\usepackage{makecell}

\usepackage{calc}
\usepackage{listings}
\definecolor{cogreen}{RGB}{0,80,0}
\def\hlRaise{1.2pt}
\makeatletter
\newenvironment{btHighlightA}[1][]
{\begingroup\tikzset{bt@Highlight@par/.style={#1}}\begin{lrbox}{\@tempboxa}}
{\end{lrbox}\bt@HLA@box[bt@Highlight@par]{\@tempboxa}\endgroup}
\newcommand\btHLA[1][]{%
  \begin{btHighlightA}[#1]\bgroup\aftergroup\bt@HLA@endenv%
}
\def\bt@HLA@endenv{%
  \end{btHighlightA}%
  \egroup
}
\newcommand{\bt@HLA@box}[2][]{%
  \tikz[#1]{%
    \pgfpathrectangle{\pgfpoint{1pt}{0pt}}{\pgfpoint{\wd #2}{\ht #2}}%
    \pgfusepath{use as bounding box}%
    \node[anchor=base west, fill=green!30,outer sep=0pt,inner xsep=0pt, inner ysep=0pt, minimum height=\ht\strutbox,#1]{\raisebox{\hlRaise}{\strut}\strut\usebox{#2}};
  }%
}
\makeatother
\makeatletter
\newenvironment{btHighlightB}[1][]
{\begingroup\tikzset{bt@Highlight@par/.style={#1}}\begin{lrbox}{\@tempboxa}}
{\end{lrbox}\bt@HLB@box[bt@Highlight@par]{\@tempboxa}\endgroup}
\newcommand\btHLB[1][]{%
  \begin{btHighlightB}[#1]\bgroup\aftergroup\bt@HLB@endenv%
}
\def\bt@HLB@endenv{%
  \end{btHighlightB}%
  \egroup
}
\newcommand{\bt@HLB@box}[2][]{%
  \tikz[#1]{%
    \pgfpathrectangle{\pgfpoint{1pt}{0pt}}{\pgfpoint{\wd #2}{\ht #2}}%
    \pgfusepath{use as bounding box}%
    \node[anchor=base west, fill=red!30,outer sep=0pt,inner xsep=0pt, inner ysep=0pt, minimum height=\ht\strutbox,#1]{\raisebox{\hlRaise}{\strut}\strut\usebox{#2}};
  }%
}
\makeatother
\makeatletter
\newenvironment{btHighlightC}[1][]
{\begingroup\tikzset{bt@Highlight@par/.style={#1}}\begin{lrbox}{\@tempboxa}}
{\end{lrbox}\bt@HLC@box[bt@Highlight@par]{\@tempboxa}\endgroup}
\newcommand\btHLC[1][]{%
  \begin{btHighlightC}[#1]\bgroup\aftergroup\bt@HLC@endenv%
}
\def\bt@HLC@endenv{%
  \end{btHighlightC}%
  \egroup
}
\newcommand{\bt@HLC@box}[2][]{%
  \tikz[#1]{%
    \pgfpathrectangle{\pgfpoint{1pt}{0pt}}{\pgfpoint{\wd #2}{\ht #2}}%
    \pgfusepath{use as bounding box}%
    \node[anchor=base west, draw=black!50,thick,outer sep=1pt,inner xsep=0pt, inner ysep=0pt, rounded corners=3pt, minimum height=\ht\strutbox+1pt,#1]{\raisebox{1pt}{\strut}\strut\usebox{#2}};
  }%
}
\makeatother
\makeatletter
\newenvironment{btHighlightD}[1][]
{\begingroup\tikzset{bt@Highlight@par/.style={#1}}\begin{lrbox}{\@tempboxa}}
{\end{lrbox}\bt@HLD@box[bt@Highlight@par]{\@tempboxa}\endgroup}
\newcommand\btHLD[1][]{%
  \begin{btHighlightD}[#1]\bgroup\aftergroup\bt@HLD@endenv%
}
\def\bt@HLD@endenv{%
  \end{btHighlightD}%
  \egroup
}
\newcommand{\bt@HLD@box}[2][]{%
  \tikz[#1]{%
    \pgfpathrectangle{\pgfpoint{1pt}{0pt}}{\pgfpoint{\wd #2}{\ht #2}}%
    \pgfusepath{use as bounding box}%
    \node[anchor=base west, fill=blue!30,outer sep=0pt,inner xsep=0pt, inner ysep=0pt, minimum height=\ht\strutbox,#1]{\raisebox{\hlRaise}{\strut}\strut\usebox{#2}};
  }%
}
\makeatother
\makeatletter
\newenvironment{btHighlightE}[1][]
{\begingroup\tikzset{bt@Highlight@par/.style={#1}}\begin{lrbox}{\@tempboxa}}
{\end{lrbox}\bt@HLE@box[bt@Highlight@par]{\@tempboxa}\endgroup}
\newcommand\btHLE[1][]{%
  \begin{btHighlightE}[#1]\bgroup\aftergroup\bt@HLE@endenv%
}
\def\bt@HLE@endenv{%
  \end{btHighlightE}%
  \egroup
}
\newcommand{\bt@HLE@box}[2][]{%
  \tikz[#1]{%
    \pgfpathrectangle{\pgfpoint{1pt}{0pt}}{\pgfpoint{\wd #2}{\ht #2}}%
    \pgfusepath{use as bounding box}%
    \node[anchor=base west, fill=orange!30,outer sep=0pt,inner xsep=0pt, inner ysep=0pt, minimum height=\ht\strutbox,#1]{\raisebox{\hlRaise}{\strut}\strut\usebox{#2}};
  }%
}
\makeatother
\lstnewenvironment{bigcode}{
    \lstset{
        basicstyle=\fontsize{8}{9}\ttfamily,
        numbers=left,
        frame=none
    }
}{}
\lstnewenvironment{bigcodeEdit}{
    \lstset{
        basicstyle=\fontsize{7}{8}\selectfont\ttfamily,
        numbers=none
    }
}{}

\usepackage{soul}
\definecolor{code-gray}{gray}{0.9}
\newcommand{\code}[1]{%
    \begingroup%
    \small%
    \texttt{#1}%
    \endgroup%
}

\newcommand{\urlx}[1]{{\href{#1}{#1}}}

\newcommand{\getafix}{Getafix}

\newcommand{\editarrow}{\rightarrowtail}
\newcommand{\editx}[2]{{#1} \ensuremath{\editarrow} {#2}}
\newcommand{\edit}[2]{\editx{\code{#1}}{\code{#2}}}

\newcommand{\predicate}[1]{\ensuremath{\textsf{#1}}\xspace}
\newcommand{\setfont}{\textsf}
\newcommand{\flagMod}{\texttt{mod}\xspace}
\newcommand{\flagUnmod}{\texttt{unmod}\xspace}
\newcommand{\setMod}{\ensuremath{\{ \flagMod, \flagUnmod \}}}

\newcommand{\powerset}{\ensuremath{\mathcal{P}}}
\newcommand{\funcarrow}{\longrightarrow}

\newcommand{\beforeFirstUse}{\textit{before}\xspace}
\newcommand{\afterFirstUse}{\textit{after}\xspace}
\newcommand{\before}{\textit{before}\xspace}
\newcommand{\after}{\textit{after}\xspace}
\newcommand{\tsub}[1]{\ensuremath{_{#1}}}
\newcommand{\tsup}[1]{\ensuremath{^{#1}}}
\newcommand{\ifnonempty}[2]{\if\relax\detokenize{#1}\relax\else{#2}\fi}
\newif\ifboldmode
\boldmodefalse
\newcommand{\node}[3]{
\ifboldmode
  \textbf{\ensuremath{\predicate{\ifnonempty{#2}{\texttt{#2}\;:\;}#1}\ifnonempty{#3}{\left(#3\right)}}}
\else
  \ensuremath{\predicate{\ifnonempty{#2}{\texttt{#2}\;:\;}#1}\ifnonempty{#3}{\left(#3\right)}}
\fi
}
\newcommand{\unode}[3]{
  \normalfont \ensuremath{\predicate{\ifnonempty{#2}{\texttt{#2}\;:\;}#1}\ifnonempty{#3}{\left(#3\right)}}
}
\newcommand{\hole}[2]{
\ifboldmode
  \node{#1}{\it \textbf h\tsup {#2}}{}
\else
  \node{#1}{\it h\tsup {#2}}{}
\fi
}
\newcommand{\uhole}[2]{\unode{#1}{\it h\tsup {#2}}{}}

\newcommand{\setHole}{\setfont{Hole}\xspace}
\newcommand{\setTree}{\setfont{Tree}\xspace}
\newcommand{\setEdit}{\setfont{Edit}\xspace}
\newcommand{\setMapping}{\setfont{Mapping}\xspace}
\newcommand{\setTreeEx}{\setfont{Tree}\tsup{+}\xspace}
\newcommand{\setEditEx}{\setfont{Edit}\tsup{+}\xspace}

\newcommand{\setLabel}{\setfont{Label}\xspace}
\newcommand{\setValue}{\setfont{Value}\xspace}
\newcommand{\setLoc}{\setfont{Location}\xspace}
\newcommand{\setTreeRef}{\setfont{TreeRef}\xspace}

\newcommand{\antiunifyName}{\predicate{antiUnify}}
\newcommand{\antiunify}[2]{\ensuremath{\antiunifyName(\text{#1}, \text{#2})}}
\newcommand{\strunchName}{\predicate{stripUnmod}}
\newcommand{\strunch}[1]{\ensuremath{\strunchName({#1})}}
\newcommand{\popunchName}{\predicate{populate}}
\newcommand{\popunch}[1]{\ensuremath{\popunchName({#1})}}
\newcommand{\mpt}{\ensuremath{\sqsubseteq}}

\usepackage[mathscr]{euscript}
\newcommand{\predMatch}{\predicate{match}}

\newcommand{\predHFix}{\predicate{hfix}}

\newcommand{\setNats}{\mathbb{N}}

\newcommand{\bigProb}[1]{\mathbb{P}\big({#1}\big)}

\definecolor{mpcolor}{rgb}{0.1,0.9,0.1}

\newtoggle{isdoubleblind}
\newcommand{\doubleblind}[2]{\iftoggle{isdoubleblind}{#1}{#2}}

\begin{document}

\title{\getafix: Learning to Fix Bugs Automatically}         


\author{Johannes Bader}
\orcid{0000-0003-3129-3814}             
\affiliation{
  \institution{Facebook}           
}
\email{jobader@fb.com}         

\author{Andrew Scott}
\orcid{nnnn-nnnn-nnnn-nnnn}             
\affiliation{
  \institution{Facebook}            
}
\email{andrewscott@fb.com}          

\author{Michael Pradel}
\orcid{nnnn-nnnn-nnnn-nnnn}             
\affiliation{
  \institution{Facebook}           
}
\email{michael@binaervarianz.de}         

\author{Satish Chandra}
\orcid{nnnn-nnnn-nnnn-nnnn}             
\affiliation{
  \institution{Facebook}           
}
\email{satch@fb.com}         

\begin{abstract}
Static analyzers help find bugs early by warning about recurring bug categories.
While fixing these bugs still remains a mostly manual task in practice, we observe that fixes for a specific bug category often are repetitive.
This paper addresses the problem of automatically fixing instances of common bugs by learning from past fixes.
We present Getafix, an approach that produces human-like fixes while being fast enough to suggest fixes in time
proportional to the amount of time needed to obtain static analysis results in the first place.

Getafix is based on a novel hierarchical clustering algorithm that summarizes fix patterns into a hierarchy ranging from general to specific patterns.
Instead of a computationally expensive exploration of a potentially large space of candidate fixes, Getafix uses a simple yet effective ranking technique that uses the context of a code change to select the most appropriate fix for a given bug.

Our evaluation applies Getafix to 1,268 bug fixes for six bug categories reported by popular static analyzers for Java, including null dereferences, incorrect API calls, and misuses of particular language constructs.
The approach predicts exactly the human-written fix as the top-most suggestion between 12\% and 91\% of the time, depending on the bug category.  The top-5 suggestions contain fixes for 526 of the 1,268 bugs.
Moreover, we report on deploying the approach within \doubleblind{a major software company}{Facebook}, where it contributes to the reliability of software used by billions of people.
To the best of our knowledge, Getafix is the first industrially-deployed automated bug-fixing tool that learns fix patterns from past, human-written fixes to produce human-like fixes.
\end{abstract}



\maketitle

\section{Introduction}\label{sec:introduction}

Modern production code bases are extremely complex and are updated constantly.
Static analyzers can help developers find potential issues (referred to as bugs for the rest of this paper) in their code, which is necessary to keep code quality high in these large code bases.
While finding bugs early via static analysis is helpful, the problem of fixing these bugs still remains a mostly manual task in practice, hampering the adoption of static analysis tools~\cite{ChristakisB16}.

Most static analyzers look for instances of common bug categories, such as potential null dereferences, incorrect usages of popular APIs, or misuses of particular language constructs.
We observe that fixes for a specific bug category often resemble each other: there is a pattern to them.
That is, past human fixes of the same bug category may offer insights into how future instances of the bug category should be fixed.
Given this observation, can we automate fixing bugs identified by learning from past fixes?

This paper addresses the problem of automatically fixing instances of common bug categories by learning from past fixes.  We assume two inputs:
(1) A set of changes that fix a specific kind of bug, e.g., from the version history of a code base.
These changes serve as training data to learn fix patterns from.
(2) A piece of code with a static analysis warning that we want to fix.
Given only these two inputs, the problem is to predict a fix that addresses the static analysis warning in a way similar or equal to what a human developer would do.
By automating the fix generation and leaving to the human only the final decision of whether to apply the fix, the overall effort spent on addressing bugs pointed out by static analyzers can be greatly reduced.

We focus on kinds of bugs that have non-trivial yet repetitive fixes.
On the one end of the spectrum, there are bug categories that trivially imply a specific fix.
For example, for a warning that suggests a field to be \code{final}, implementing an automated fix suggestion is straightforward. Such an auto-fix can be defined by the author of that rule in the static analyzer, without knowing the specific context in which the rule is applied; indeed, some of Error Prone~\cite{Aftandilian2012} rules come with auto-fixes.
On the other end of the spectrum are bugs that require complex, application-specific fixes, such as an issue with a UI tab not displaying after a specific series of interactions from a user.
Here we target bug categories in between these two extremes, where finding a fix is non-trivial, yet typical fixes belong to a set of recurring fix patterns.
For such bug categories, there often exists more than one way to fix the problem, and the ``right'' way to address a specific instance of the bug category depends on the context, e.g., the code surrounding the static analysis warning.

\begin{figure*}
	\begin{minipage}[t]{0.3\linewidth}
\textit{Extend existing if condition conjunctively:}
\begin{bigcodeEdit}[numbers=none]
// ...
if (key != 0 +++␣&& mThread != null+++) {
  mThread.cancel(key);
} else {
  ThreadPool.tryCancel(key);
}
// ...
\end{bigcodeEdit}
	\end{minipage}
	\hspace{1em}
	\begin{minipage}[t]{0.3\linewidth}
\textit{Add an early return:}
\begin{bigcodeEdit}
void updateView() {
  // ...
  View v = ViewUtil.findView(name);
  +++if (v == null) {+++
  +++  return;+++
  +++}+++
  data.send(((UpdatableView) v)
      .getContentMgt(), "UPDATE");
  // ...
}
\end{bigcodeEdit}
	\end{minipage}
	\hspace{1em}
	\begin{minipage}[t]{0.3\linewidth}
\textit{Protect call with conditional expression:}
\begin{bigcodeEdit}
@Override
public void onClose() {
  final Window w =
  +++win != null ? +++win.get()+++ : null+++;
  if (w != null
      && w.getProc().isActive()) {
    w.getProc().removeListeners();
  }
}
\end{bigcodeEdit}
	\end{minipage}
	\caption{Different real-world fixes of potential null dereference bugs.}
	\label{fig:NPEexamples}
\end{figure*}

As an example of a bug category targeted in this work, consider \code{NullPointerException}s -- still one of the most prevailing bugs in Java and other languages.
If a static analyzer warns about a potential null dereference, developers may fix the problem in various ways.
Figure~\ref{fig:NPEexamples} shows three anonymized examples of fixes of null dereference bugs, which add a conjunct to an existing if condition, replace a call with a ternary operation, and add an early return, respectively.
While all these fixes introduce some kind of null check, the exact fix heavily depends on the already existing code.
Beyond these examples, there are even more ways of fixing null dereference bugs, e.g., by adding a new if statement or by extending an existing if condition disjunctively.
Learning all these fix patterns and deciding which one to apply to a given piece of buggy code is a non-trivial problem.

Our work aims at automating bug fixing in large-scale, industrial software development.
This setting leads to several interesting challenges to deal with:
\begin{itemize}
	\item To reduce the human time spent on fixing a bug, the approach may propose only a \emph{small number} of potential fixes, ideally only one fix.
	\item To make this fix acceptable to developers, the suggested fix should be \emph{human-like}: very similar to or exactly the same as a fix a human developer would implement.
	\item To suggest fixes \emph{quickly}, as well as to keep the computing resources required to find fixes within bounds, the approach cannot explore a large space of candidate fixes and validate each of them against a test suite or any other computationally expensive validation routine.

\end{itemize}

Addressing these challenges, we present Getafix, an automated technique that learns recurring fix patterns for static analysis warnings and suggests fixes for future occurrences of the same bug category.
Getafix produces \emph{human-like} fixes, and does so \emph{fast enough} (typically within 10 seconds) to offer a fix suggestion in roughly the
same magnitude of time as a human developer waits for static analysis results anyway.
In a nutshell, the approach consists of three main steps.
First, it splits a given set of example fixes into AST-level edit steps.
Second, it learns recurring fix patterns from these edit steps, based on a novel hierarchical, agglomerative clustering technique that produces a hierarchy of fix patterns ranging from very general to very specific fixes.
Third, given a previously unseen bug to fix, Getafix finds suitable fix patterns, ranks all candidate fixes, and suggests the top-most fix(es) to the developer.
As a check during the third step, Getafix validates each suggested fix against the static analyzer to ensure that the fix removes the warning.
Note that the validation against the static analyzer is a one-time effort per fix, keeping the computational resources within reasonable bounds.

The Getafix approach and the constraints that motivated it differ from the assumptions made by automated program repair techniques~\cite{cacm2019-program-repair}, the most practical of which for large systems are
\textit{generate-and-validate} repair systems ~\cite{Goues2012GenProgAG,Kim2013Automatic,Le2016HistoryDP}.

\begin{enumerate}

\item
Because it must produce fixes quickly, Getafix neither generates, nor validates a large number of fix candidates.  Getafix produces a ranked list of fix candidates based entirely on past fixes it knows about and on the context in which a fix is applied.  This is done without \emph{any} validation; Getafix's ranking reflects the confidence it has in the fix. It then offers to the developer one (or a few, configurable) top-ranked fix(es) after a validation step. Note that Getafix does not go down its list exploring additional fix candidates until a validated fix is found.

\item
In contrast to generate-and-validate repair systems that use a test suite as validation filter, Getafix uses the same static analyzer as the validation filter to make sure the warning goes away.  This is because at this point
in the development, we cannot assume that the code is ready to pass a test suite.

%

\item Prior repair techniques rely on statistical fault localization (e.g., based on coverage of passing and failing test cases) that determines where in the code a bug should be fixed.  Getafix avoids this step by exploiting the fact that static analyzers pinpoint
the location of the bug.
Consequently, it also avoids the need to try applying fix candidates at various potential fault locations.

\item
Finally, the goal that Getafix sets for itself is not merely a fix that makes its validation pass, but is
as close as possible to what a human would do in that situation. (While a generic null check would suppress a
null dereference warning, as Figure~\ref{fig:NPEexamples} shows, a human might choose a very specific kind of fix in a particular context.)
By contrast, most program repair systems aim to produce a fix that is functionally correct, but may not match a human fix.

\end{enumerate}

Another related line of work is on learning edit patterns, including potential bug fix patterns, from version histories~\cite{Rolim2017LearningSPT,Rolim2018LearningQF,Brown2017a}.
In contrast to Getafix, these techniques learn from \emph{all} code changes, leaving the task of identifying interesting fix patterns to a human.
A key insight of our work is that learning from fixes for a specific static analysis warning avoids this human effort.

We evaluate Getafix in two ways.
One part of our evaluation applies the approach to a total of 1,268 bug fixes for six kinds of warnings reported by two publicly available and widely used static analyzers for Java.
The bug categories include potential null dereferences, incorrect uses of Java's reference equality, and common API misuses.
After learning fix patterns from several dozens to several hundreds of examples, Getafix predicts exactly the human fix as the top-most suggestion for 12\% to 91\% of all fixes, depending on the bug category.
In a setting where developers are willing to inspect up to five fix suggestions, the percentage of correctly predicted fixes even ranges between 19\% and 92\%, containing fixes for 526 of the 1,268 bugs.
Because these results indicate how often the predicted fix exactly matches the human fix, as opposed to producing any human-acceptable fix, these results provide a lower bound of the effectiveness of Getafix.

The other part of our evaluation deploys Getafix to production at \doubleblind{a major software company (henceforth called CompanyX)}{Facebook}, where it now contributes to the stability of apps used by billions of people.
At \doubleblind{CompanyX}{Facebook}, Getafix currently suggests fixes for bugs found by Infer~\cite{calcagno2015moving}\doubleblind{}{\footnote{\urlx{https://code.fb.com/developer-tools/open-sourcing-facebook-infer-identify-bugs-before-you-ship/}}}, a static analysis tool that identifies issues, such as null dereferences, in Android and Java code.
For example, the fixes in Figure~\ref{fig:NPEexamples} have been suggested by Getafix and accepted by developers at \doubleblind{CompanyX}{Facebook}.
We find that developers accept around 42\% of all fixes suggested by Getafix, helping to save precious developer time by addressing bugs with a single click.

~\\\noindent
\paragraph*{Contributions}
This paper makes the following contributions:

\begin{itemize}
	\item Fully automated suggestion of bug fixes for static analysis bug reports, computed in seconds,
	without requiring computationally expensive search over a large space of candidate bug fixes.
	\item A novel clustering technique for discovering repetitive fix patterns in a way that is more general than a previous greedy approach and that preserves important context needed for applying fixes.
	\item Simple yet effective statistical ranking technique to select the best fix(es) to suggest, which enables predicting human-like fixes among the top few suggestions.
	\item Empirical evidence that the technique works well for a diverse set of bug categories taken from Error Prone and Infer.

\end{itemize}

To the best of our knowledge, Getafix is the first industrially-deployed automated bug-fixing tool that
learns fix patterns from past, human-written commits, and produces human-like fixes in a short
amount of time.

\section{Overview}\label{sec:overview}

\begin{figure*}
	\includegraphics[width=.8\linewidth,clip,trim=1.4in 5.7in 2.1in 0.5in]{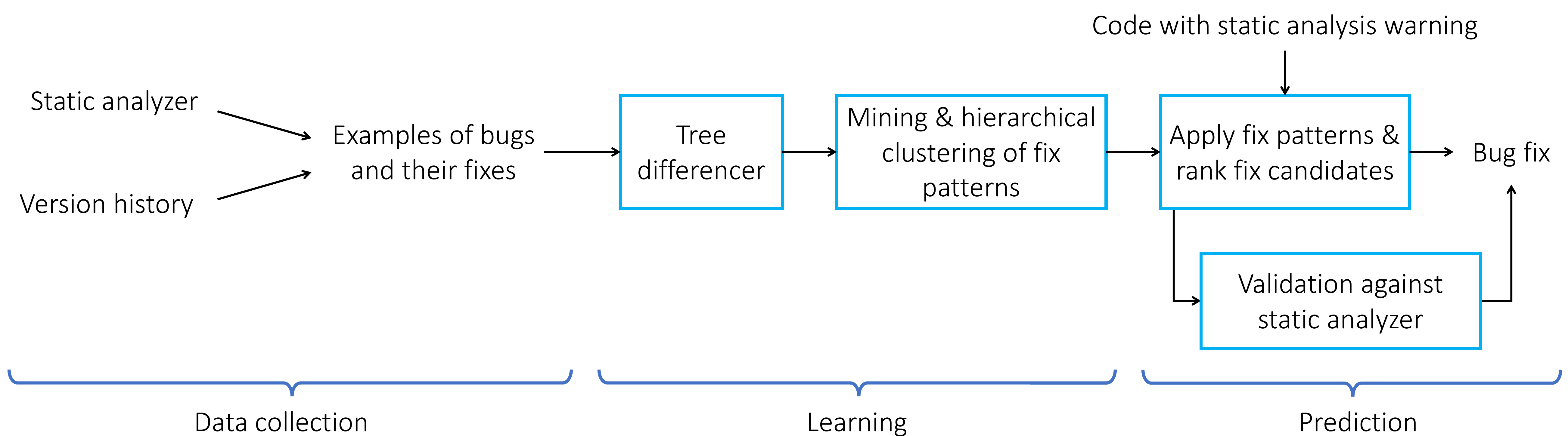}
	\caption{Overview of Getafix.}
	\label{fig:overview}
\end{figure*}

Getafix consists of three main components, organized in a learning phase and a prediction phase.
In the following we will describe their functionality and challenges at a high level, followed by a more detailed description in later sections.
Figure~\ref{fig:overview} gives an overview of the approach.
During the learning phase, a set of pairs of bugs and their fixes is given as training data to Getafix.
As training data can serve any collection of past human code changes tied to a specific signal, such as a static analysis warning, a type error, a lint message, or simply the fact that a change was suggested during human code review.
Our evaluation focuses on static analysis warnings as the signal, i.e., all bugs and fixes have been detected as instances of a specific bug category by a static analyzer, e.g., as potential null dereferences.
During the prediction phase, Getafix then takes previously unseen code that comes with the same signal as the training examples and produces a bug fix.

The first step of the learning phase is a \textit{tree differencer}, which identifies changes at the AST level.
It generates \textit{concrete edits}, which are pairs of sub-ASTs of the original \beforeFirstUse and \afterFirstUse ASTs,
representing a specific change that could be replayed on different code by replacing instances of the edit's \before AST with its \after AST.

Based on the concrete, AST-level edits, the second step is a new way of \textit{learning fix patterns}.
To generalize from specific fix examples, fix patterns have ``holes'', i.e., pattern variables, that may match specific subtrees, similar to the representation used by \citet{Rolim2017LearningSPT} and \citet{Long2017Automatic}.
A technical contribution is a novel hierarchical, agglomerative clustering technique that organizes fix patterns into a hierarchical structure.
The approach derives different variants of fix patterns, ranging from very specific patterns that match few concrete edits to very general fix patterns that match many concrete edits.
Another contribution is to include into the fix patterns not only the code changes themselves, but also some surrounding context.
This context is important to decide which out of multiple possible fix patterns to apply to a given piece of code.

After learning fix patterns, which is a once-per-bug-category effort, the prediction phase of Getafix \emph{applies the patterns} to previously unseen buggy code to produce a suitable fix.
Since Getafix aims at predicting fixes without a computationally expensive validation of many fix candidates, it internally \emph{ranks candidate fixes}.
We present a simple yet effective, statistical ranking technique that uses the additional context extracted along with fix patterns.

Before suggesting a fix to the developer, Getafix validates the predicted fix against the same tool that provided the signal for the bug category, e.g., a static analysis, type checker, or linter.
Getafix is agnostic to the signal used, so we assume a possibly computationally expensive black box component.

\section{Tree differencer}\label{sec:treediff}

\begin{figure}[tb]

	\def\editColWidth{\dimexpr.2\textwidth}
	\hspace{0pt}\\*\noindent
	\begin{tabular}{@{}p{\editColWidth}@{}p{\dimexpr.05\textwidth}@{}p{\editColWidth}@{}}
		\begin{minipage}{\editColWidth}
			{\begin{bigcodeEdit}
public class Worker {
  ***public void doWork() {***
    ***task.makeProgress();***
  ***}***
  ~~~public~~~long getRuntime() {
    return now - start;
  }
}
			\end{bigcodeEdit}}
		\end{minipage}&
		\centering$\editarrow$&
		\begin{minipage}{\editColWidth}
			{\begin{bigcodeEdit}
public class Worker {
  ~~~private~~~long getRuntime() {
    return now - start;
  }
  ***public void doWork() {***
    +++if (task == null)+++
      +++return;+++
    ***task.makeProgress();***
  ***}***
}
			\end{bigcodeEdit}}
		\end{minipage}
	\end{tabular}

	\caption{Concrete edits extracted by the tree differencer.}
	\label{fig:sample_treediff}
\end{figure}

The first step of Getafix takes a set of example bug fixes, and decomposes each fix into fine-grained edits.
These edits provide the basic ingredients for learning and applying fix patterns in later steps.
For example, Figure \ref{fig:sample_treediff} shows a code change and the three fine-grained edits that Getafix extracts: (i) inserting an early return if \code{task} is \code{null} (shown in green); (ii) changing \code{public} to \code{private} (shown in red); and (iii) moving the \code{doWork} method (shown in blue).

\subsection{Trees}

Getafix extracts fine-grained edits based on ASTs.
For our purposes, a node in an AST has:
\begin{itemize}
	\item a \textit{label}, e.g., "BinEx" (for binary expression), "Literal", or "Name",
	\item a possibly empty \textit{value}, e.g., \code{+}, \code{42}, or \code{foo}, and
	\item \textit{child nodes}, each having a \textit{location} that describes their relationship to the parent node, e.g., "left" and "right" to address the left/right subexpression of a binary expression.
\end{itemize}
More formally, we define the set of trees as follows.
\begin{definition}[Tree]\label{def:tree}
	\begin{align*}
		\setTree &= \setLabel \times \setValue \times \bigcup_{k \in \setNats} (\setLoc \times \setTree)^k
	\end{align*}
	where \setLabel, \setLoc and \setValue are sets of strings.
\end{definition}

For readability, we represent ASTs using term notation rather than tuples.
For example, parsing \code{x = y + 2} results in an AST \unode{Assign}{}{\unode{Name}{x}{}, \unode{BinEx}{+}{\unode{Name}{y}{}, \unode{Literal}{2}{}}}.
Child nodes are listed in parentheses following their parent node.
Parentheses are omitted if no child nodes exist.
Values, if not empty, are prepended to labels using syntax borrowed from type judgments.

\subsection{Tree Edits}

Given two trees, we define edits as follows.

\begin{definition}[Tree edits]\label{def:edit}
	\begin{align*}
		\setEdit &= \setTree \times \setTree \times \powerset^\setMapping\\
		\setMapping &= \setTreeRef \times \setTreeRef \times \setMod
	\end{align*}
\end{definition}

Edits are triples containing the \before and \after ASTs, followed by a set of mappings.
A mapping is a triple referencing a node of the \before and a node of the \after AST, as well as a flag indicating whether the pair of subtrees is part of a modification (\flagMod means the node is part a modification, \flagUnmod means that the mapped subtree is unmodified).
\setTreeRef is the set of references uniquely identifying a node within an AST.

\boldmodetrue
We write edits in source code as \edit{\before code}{\after code}, e.g., \edit{x = y + 2}{x = 3 + y}.

Combining this notation with our AST notation from above, we write tree edits as \edit{\before AST}{\after AST}, where we write modified subtrees bold (whenever the distinction is relevant) and give nodes connected by a mapping matching indices.
For example, two concrete edits emitted for the above code change are
\editx{\node{BinEx\tsub 0}{+}{\node{Name\tsub 1}{y}{}, \node{Literal}{2}{}}}{\node{BinEx\tsub 0}{+}{\node{Literal}{3}{}, \node{Name\tsub 1}{y}{}}}
and\\
\editx{\node{Assign\tsub 0}{}{\unode{Name\tsub 1}{x}{}, \node{BinEx\tsub 2}{+}{\node{Name\tsub 3}{y}{}, \node{Literal}{2}{}}} \linebreak}{\node{Assign\tsub 0}{}{\unode{Name\tsub 1}{x}{}, \node{BinEx\tsub 2}{+}{\node{Literal}{3}{}, \node{Name\tsub 3}{y}{}}}}.
\boldmodefalse


An alternative to tree-based reasoning about edits would be to use a line-based diffing tool.
However, line-based diffing reasons about code on a more coarse-grained level, ignoring the structural information provided by ASTs.
For example, given the change in Figure~\ref{fig:sample_treediff}, line-based diffing would mark both methods as fully removed and inserted, whereas tree-based edits can encode the move and hence also detect the insertion within the moved method as a concrete edit.

\subsection{Extracting Concrete Edits}

To extract edits from a given pair of ASTs, Getafix builds on the GumTree algorithm~\cite{Falleri2014FineGrained}, a tree-based technique to compute fine-grained edit steps that lead from one AST to another.
The approach extracts four kinds of edit steps: deletion, insertion, move, and update.
An unmapped node in the \before or \after tree is considered a \textit{deletion} or \textit{insertion}, respectively.
A mapped pair of nodes whose parents are not mapped to each other is considered a \textit{move}.
A mapped pair of nodes with different values is considered an \textit{update} and may be part of a move at the same time.
Any pairs of nodes involved in one of the above operations is consider to be \emph{modified}, whereas all other pairs of mapped nodes are considered \textit{unmodified}.
An entire subtree is considered \textit{unmodified} if all its nodes are unmodified.


For the example in Figure \ref{fig:sample_treediff}, the insertion of the if statement is an addition, the change from \code{public} to \code{private} is an update, and method \code{doWork} has been moved.
The subtree representing the call \code{task.makeProgress()} is unmodified.
In contrast, the subtree representing the block surrounding this call is modified, due to the insertion of the if statement.

As demonstrated with the insertion combined with a move in Figure \ref{fig:sample_treediff}, modifications can be nested and it is unclear what the granularity of a concrete edit should be.
Maybe the insertion itself is the fix, maybe the move is important.
Our algorithms are language agnostic, so without domain knowledge there is no robust strategy for grouping modifications into concrete edits.
We address this challenge by extracting an entire \textit{spectrum} of concrete edits based on the modifications reported by GumTree:
Any pair of mapped nodes will be turned into the roots of a new concrete edit, if that concrete edit contains at least one modification.
In the example of Figure~\ref{fig:sample_treediff}, we would hence extract concrete edits rooted at the body of \code{doWork} (contains the insertion), at the method declaration of \code{doWork} (contains the insertion as well), at the method declaration of \code{getRuntime} (contains the update), at the class body level (contains all modifications), and any of its ancestors for the same reason.
By contrast, we will not emit a concrete edit rooted at the body of \code{getRuntime} as it does not contain a modification.
In addition to making concrete edits rooted at mapped nodes, we also create a concrete edit rooted at the parent of each contiguous region of modified nodes in the before and after trees.
This is done to ensure that we still learn patterns from edits which occur near other changes.

The approach is designed to extract too many rather than too few concrete edits.
The rationale is that the clustering step of Getafix, which is explained in detail in the following, automatically prioritizes patterns with the best level of granularity, since there will exist multiple similar concrete edits.
For instance, when combining the concrete edits emitted for Figure \ref{fig:sample_treediff} with further concrete edits from bug fixes for null dereferences, the insertion of an if statement is likely to appear more often than the move or update.
In contrast, concrete edits containing further modifications, e.g., rooted further up in the AST, are likely to be discarded as noise, because there are no similar concrete edits to create a cluster with.

\section{Learning Fix Patterns}\label{sec:mining}

Given the set of fine-grained, tree-level edits extracted by the tree differencer described in the previous section, the next step is to learn fix patterns, i.e., recurring edit patterns observed in fixes for a specific kind of bug.
Intuitively, an edit pattern can be thought of as a generalization of multiple edits.
To abstract away details of concrete edits, edit patterns may have ``holes'', or pattern variables, to represent parts of a tree where concrete edits differ.
A key contribution of Getafix is a novel algorithm that derives a hierarchy of edit patterns, which contains edit patterns at varying levels of generality, ranging from concrete edits at the leaves to abstract edit patterns at the root.

The algorithm to derive a hierarchy of edit patterns is based on a generalization operation on edit patterns, which we obtain through anti-unification~\cite{Kutsia2014AntiUnification}, an existing method of generalizing among different symbolic expressions.
We start by presenting the generalization operation (Section~\ref{sec:anti-unification}) and then show how Getafix uses this operation to guide a hierarchical clustering algorithm (Section~\ref{sec:clustering}).
Finally, Section~\ref{sec:context} describes how Getafix augments edit patterns with context information that helps deciding when to apply a learned edit pattern to new code.

\subsection{Generalizing Edit Patterns via Anti-Unification}
\label{sec:anti-unification}

\subsubsection{Tree Patterns and Tree Edit Patterns}

A set of concrete trees can be abstracted into a tree pattern.
Formally, we extend our definition of trees by adding holes:
\begin{definition}[Tree pattern]
Let $\setHole = (\setLabel \cup \{ ? \}) \times \setNats$, then the set of tree patterns is
$$\setTreeEx = \setLabel \times \setValue \times \bigcup_{k \in \setNats} (\setLoc \times \setTreeEx)^k \cup \setHole$$
\end{definition}
Holes may have a label $\alpha$ (we write \hole{$\alpha$}{k}) in which case they match an arbitrary subtree that has a root with label $\alpha$.
If the label is omitted (we write \hole{?}{k}) the hole matches any subtree.
Holes are indexed, allowing tree patterns to express whether two holes must match identical subtrees.

For example, consider the following trees $t_1, t_2, t_3 \in \setTree$ and tree patterns $p_1, p_2, p_3, p_4 \in \setTreeEx$:
\begin{align*}
	t_1 &= \node{BinEx}{+}{\node{Name}{x}{}, \node{Literal}{42}{}}\\
	t_2 &= \node{BinEx}{+}{\node{Name}{x}{}, \node{Name}{y}{}}\\
	t_3 &= \node{BinEx}{+}{\node{Name}{x}{}, \node{Name}{x}{}}\\
	p_1 &= \node{BinEx}{+}{\hole{Name}{0}, \hole{Literal}{1}{}}\\
	p_2 &= \hole{BinEx}{0}\\
	p_3 &= \node{BinEx}{+}{\hole{Name}{0}, \hole{?}{1}}\\
	p_4 &= \node{BinEx}{+}{\hole{?}{0}, \hole{?}{0}}
\end{align*}
Pattern $p_1$ matches only $t_1$, since $t_2$ and $t_3$ do not match the label of hole {\ensuremath{h^1}}.
Patterns $p_2$ and $p_3$ match all three ASTs.
Finally, $p_4$ matches only $t_3$, since the pattern requires both operands of the binary expression to be identical.

Similar to tree patterns, we define the set \setEditEx of edit patterns analogous to \setEdit (see Definition~\ref{def:edit}), but using \setTreeEx instead of \setTree.

\subsubsection{Generalizing Tree Patterns}

To learn recurring edit patterns, Getafix generalizes concrete edits into patterns through a process called anti-unification.
As an intermediate step, we first present how to generalize two tree patterns into a generalization that describes both of them.
The generalization is constructed such that each hole has a set of substitutions that can be used to recover the original input tree patterns.
There always exists a trivial generalization, which has a single hole at the root.
However, that generalization is generally not very interesting and therefore anti-unification seeks to find the least general generalization that can describe the input trees, meaning the generalization retains as much information as possible.

\boldmodetrue
\begin{figure}
	\begin{align*}
		&\antiunify{
			\node{$\boldsymbol{\alpha}$}{\ensuremath{\boldsymbol{v}}}{\ensuremath{\boldsymbol{a_1, ...\;, a_n}}}
		}{
			\node{$\boldsymbol{\beta}$}{\ensuremath{\boldsymbol{w}}}{\ensuremath{\boldsymbol{b_1, ...\;, b_m}}}
		} =\\
		&\hspace{0.5cm}\begin{cases}
			\node{$\boldsymbol{\alpha}$}{\ensuremath{\boldsymbol{v}}}{\ensuremath{\boldsymbol{c_1, ...\;, c_n}}}
			& \text{ if }\alpha = \beta \wedge v = w \wedge n = m\\
			\hspace{0.8cm}\text{\rlap{where $c_i = \antiunify{$a_i$}{$b_i$}\;$ $\forall \;i \in \{1, ..., n\}$}}&\\
			\hole{$\boldsymbol{\alpha}$}{k}
			& \text{ if }\alpha = \beta \wedge (v \neq w \vee n \neq m)\\
			\hole{?}{k} & \text{ otherwise}
		\end{cases}\\
		&\hspace{2cm}\text{\llap{where }}n, m \in \setNats,\;\; a_1, ..., a_n, b_1, ..., b_m \in \setTreeEx\\
		&\hspace{2cm}\alpha, \beta \in \setLabel,\;\; v, w \in \setValue,\\
		&\hspace{2cm}\text{fresh hole index } k \in \setNats\\
		&\antiunify{
			\node{$\boldsymbol{\alpha}$}{\ensuremath{\boldsymbol{v}}}{\ensuremath{\boldsymbol{a_1, ...\;, a_n}}}
		}{
			\hole{$\boldsymbol{\beta}$}{t}
		} =\\
		&\antiunify{
			\hole{$\boldsymbol{\alpha}$}{s}
		}{
			\node{$\boldsymbol{\beta}$}{\ensuremath{\boldsymbol{w}}}{\ensuremath{\boldsymbol{b_1, ...\;, b_m}}}
		} =\\
		&\antiunify{
			\hole{$\boldsymbol{\alpha}$}{s}
		}{
			\hole{$\boldsymbol{\beta}$}{t}
		} =\\
		&\hspace{0.5cm}\begin{cases}
			\hole{$\boldsymbol{\alpha}$}{k}
			& \text{ if }\alpha = \beta\\
			\hole{?}{k} & \text{ otherwise}
		\end{cases}\\
		&\hspace{2cm}\text{\llap{where }}n, m, s, t \in \setNats,\;\; a_1, ..., a_n, b_1, ..., b_m \in \setTreeEx\\
		&\hspace{2cm}\alpha, \beta \in \setLabel,\;\; v, w \in \setValue,\\
		&\hspace{2cm}\text{fresh hole index } k \in \setNats\\
	\end{align*}
	\vspace{-1.5em}
	\caption{Anti-unification of tree patterns. For conciseness, we assume the \antiunifyName function performs memoization, so that it reuses holes where possible, without explicitly tracking them.}
	\label{fig:def-anti-unify-trees}
\end{figure}
\boldmodefalse

Figure \ref{fig:def-anti-unify-trees} formally describes the anti-unification function for tree patterns.
The function is defined recursively on the tree structure.
It merges subtrees that have the same labels, the same values, and the same number of children into a combined subtree, and replaces them by a hole otherwise.
Merging a subtree with a subtree represented by a hole yields another hole.
When merging two holes, the function keeps the label of the hole whenever it matches.

For example, let
$$a = \node{Assign}{}{\node{Name}{a}{}, \node{BinEx}{+}{\node{Name}{a}{}, \node{Name}{a}{}}}$$
and
$$b = \node{Assign}{}{\node{Name}{b}{}, \node{BinEx}{+}{\node{Name}{b}{}, \node{Literal}{2}{}}}$$
be the ASTs for assignment statements \code{a = a + a;} and \code{b = b + 2;}, respectively.
Then $\antiunify{$a$}{$b$} = $
$$\node{Assign}{}{\hole{Name}{0}, \node{BinEx}{+}{\hole{Name}{0}, \hole{?}{1}}}$$
Note how hole \hole{Name}{0} can be reused since in both instances it substitutes variables \code{a} and \code{b}.
While \hole{?}{1} also substitutes \code{a} in AST $a$, it does not substitute \code{b} in AST $b$, so a fresh hole is created.
Because the hole's substitutions have mismatching labels, \code{?} is used.
Instead of the above pattern, $\node{Assign}{}{\hole{Name}{0}, \node{BinEx}{+}{\hole{Name}{1}, \hole{?}{2}}}$ also generalizes $a$ and $b$, but the information that \hole{Name}{0} and \hole{Name}{1} substitute identical subtrees would be lost.
Pattern $\node{Assign}{}{\hole{?}{0}, \node{BinEx}{+}{\hole{?}{1}, \hole{?}{2}}}$ would be an even more general generalization, dropping the information about labels even though they matched.
Instead of computing these generalizations, anti-unification produces the least general generalization of the given patterns.

\subsubsection{Generalizing Edit Patterns}

We now extend the idea of generalizing tree patterns to tree edit patterns, which yields the generalization operation at the core of Getafix's learning algorithm.
Figure \ref{fig:def-anti-unify-edits} formally defines anti-unification of edit patterns.
The basic idea is to first anti-unify the \before trees and then the \after trees, while using a single set of substitutions between the \before and \after trees, to indicate that a hole in the \before tree corresponds to the same AST node in the \after tree.
To focus the generalized edit patterns on the modified parts of a tree, the approach uses a helper function \strunchName to drop unmodified subtrees before anti-unifying the trees.
Afterwards, Getafix populates the unmodified nodes back and maps them to each other.

\boldmodetrue
\begin{figure*}
	\begin{align*}
		&\antiunify{\editx{$\mathit{before}_1$}{$\mathit{after}_1$}}{\editx{$\mathit{before}_2$}{$\mathit{after}_2$}} = \popunch{\editx{\mathit{before}_g}{\mathit{after}_g}, \editx{\mathit{before}_1}{\mathit{after}_1}, \editx{\mathit{before}_2}{\mathit{after}_2}} \\
		&\text{where}\\
		&\hspace{0.5cm}\mathit{before}_g = \antiunify{\strunch{\mathit{before}_1}}{\strunch{\mathit{before}_2}} \\
		&\hspace{0.5cm}\mathit{after}_g = \antiunify{\strunch{\mathit{after}_1}}{\strunch{\mathit{after}_2}}\\
		&\hspace{1.5cm}\text{\llap{for }} \editx{\mathit{before}_1}{\mathit{after}_1}, \editx{\mathit{before}_2}{\mathit{after}_2} \in \setEditEx\\
		&~\\[-0.3cm]
		&\hspace{0.5cm}\strunch{\node{$\boldsymbol{\alpha}$}{\ensuremath{\boldsymbol{v}}}{\boldsymbol{\overline{a_{1..i-1}}\;,\;} a_i \boldsymbol{\;, \overline{a_{i+1..n}}}}} =
			\strunch{\node{$\boldsymbol{\alpha}$}{\ensuremath{\boldsymbol{v}}}{\boldsymbol{\overline{a_{1..i-1}}, \overline{a_{i+1..n}}}}}\\
		&\hspace{1cm}\text{\footnotesize (drop unmodified children)}\\
		&\hspace{0.5cm}\strunch{\node{$\boldsymbol{\alpha}$}{\ensuremath{\boldsymbol{v}}}{\boldsymbol{a_1, ...\;, a_n}}} = \node{$\boldsymbol{\alpha}$}{\ensuremath{\boldsymbol{v}}}{\boldsymbol{\strunch{\ensuremath{\boldsymbol{a_1}}}, ...\;, \strunch{\ensuremath{\boldsymbol{a_n}}}}}\\
		&\hspace{1cm}\text{\footnotesize (if no more children are unmodified, recurse)}\\
		&\hspace{1.5cm}\text{\llap{for }}n, i \in \setNats,\;\; a_1, ..., a_n \in \setTreeEx,\;\; \alpha \in \setLabel,\;\; v \in \setValue\\
		&~\\[-0.3cm]
		&\hspace{0.5cm}\popunch{e_g, e_1, e_2} = \text{Add back unmodified nodes and mappings}
	\end{align*}
	\vspace{-.5em}
	\caption{Anti-unification of edit patterns. To reuse holes created for the \before and \after trees, we assume that \antiunifyName uses memoization across calls to the function.}
	\label{fig:def-anti-unify-edits}
\end{figure*}

As an example, assume the training data contains two source edits:\\
\edit{\{ f(); g(); x = 1; \}}{\{ f(); x = 1; if (c) g(); \}}\\
and\\
\edit{\{ return; y = 2; \}}{\{ y = 2; if (c) onResult(); \}}.
The corresponding concrete edits are the following:
\begin{align*}
	& \mathit{before}_1 = \node{Block\tsub 0}{}{}(\unode{Call\tsub 1}{f}{},\\
	&\hspace{2.6cm}\node{Call\tsub 2}{g}{},\\
	&\hspace{2.6cm}\unode{Assign\tsub 3}{}{\unode{Name\tsub 4}{x}{}, \unode{Num\tsub 5}{1}{}})\\
	& \mathit{after}_1 = \node{Block\tsub 0}{}{}(\unode{Call\tsub 1}{f}{},\\
	&\hspace{2.4cm}\unode{Assign\tsub 3}{}{\unode{Name\tsub 4}{x}{}, \unode{Num\tsub 5}{1}{}},\\
	&\hspace{2.4cm}\node{If}{}{\node{Name}{c}{}, \node{Call\tsub 2}{g}{}})\\
	& \mathit{before}_2 = \node{Block\tsub 6}{}{}(\node{Return}{}{},\\
	&\hspace{2.6cm}\unode{Assign\tsub 7}{}{\unode{Name\tsub 8}{y}{}, \unode{Num\tsub 9}{2}{}})\\
	& \mathit{after}_2 = \node{Block\tsub 6}{}{}(\unode{Assign\tsub 7}{}{\unode{Name\tsub 8}{y}{}, \unode{Num\tsub 9}{2}{}},\\
	&\hspace{2.4cm}\node{If}{}{\node{Name}{c}{}, \node{Call}{onResult}{}})\\
\end{align*}
We first drop the unmodified subtrees and anti-unify the result:
{\begingroup\makeatletter\def\f@size{9.5}\check@mathfonts
	\begin{align*}
		&\strunch{\mathit{before}_1} = \node{Block\tsub 0}{}{\node{Call\tsub 2}{g}{}}\\
		&\strunch{\mathit{after}_1} = \node{Block\tsub 0}{}{\node{If}{}{\node{Name}{c}{}, \node{Call\tsub 2}{g}{}}}\\
		&\strunch{\mathit{before}_2} = \node{Block\tsub 6}{}{\node{Return}{}{}}\\
		&\strunch{\mathit{after}_2} = \node{Block\tsub 6}{}{\node{If}{}{\node{Name}{c}{}, \node{Call}{onResult}{}}}
	\end{align*}
\begin{align*}
	\mathit{before}_g &= \antiunify{\strunch{\mathit{before}_1}}{\strunch{\mathit{before}_2}}\\
	&= \node{Block}{}{\hole{?}{0}}\\
	\mathit{after}_g &= \antiunify{\strunch{\mathit{after}_1}}{\strunch{\mathit{after}_2}}\\
	&= \node{Block}{}{\node{If}{}{\node{Name}{c}{}, \hole{Call}{1}}}
\end{align*}
\endgroup}

We then attempt to populate back mappings where possible.
Mappings are re-established between any pair of nodes that are the result of anti-unifying mapped nodes.
Here, both edits agree about the blocks mapping to each other (mappings $0$ and $6$):
\begin{align*}
	&\mathit{before}_g' = \node{Block\tsub {10}}{}{\hole{?}{0}}\\
	&\mathit{after}_g' = \node{Block\tsub {10}}{}{\node{If}{}{\node{Name}{c}{}, \hole{Call}{1}}}
\end{align*}
Finally, Getafix populates back unmodified nodes where possible. Both edits have an assignment statement, \unode{Assign\tsub{11}}, that swaps places with the modified statement.
The approach anti-unifies this unmodified node and adds it back, giving the final edit pattern \edit{$\mathit{before}_g''$}{$\mathit{after}_g''$}:
\begin{align*}
	&\mathit{before}_g'' = \node{Block\tsub {10}}{}{}(\hole{?}{0},\\
	&\hspace{2.8cm}\unode{Assign\tsub {11}}{}{\uhole{Name}{2}, \uhole{Num}{3}})\\
	&\mathit{after}_g'' = \node{Block\tsub {10}}{}{}(\unode{Assign\tsub {11}}{}{\uhole{Name}{2}, \uhole{Num}{3}},\\
	&\hspace{2.6cm}\node{If}{}{\node{Name}{c}{}, \hole{Call}{1}})
\end{align*}
\boldmodefalse


\subsection{Hierarchical Clustering of Edit Patterns}
\label{sec:clustering}

Based on the generalization operation described above, Getafix uses a hierarchical clustering algorithm to derive a hierarchy of recurring edit patterns.
The basic idea is to start with all concrete edits and to iteratively combine pairs of edits until all edits are combined into a single edit pattern.
At each iteration, the algorithm picks a pair of edit patterns to combine, so that their anti-unification yields the least loss of concrete information.
The sequence of generalization steps eventually yields a hierarchy of edit patterns, where leaf nodes represent concrete edits and higher-level nodes represent increasingly generalized edit patterns.

\begin{figure*}[tb]
	\includegraphics[width=0.7\textwidth]{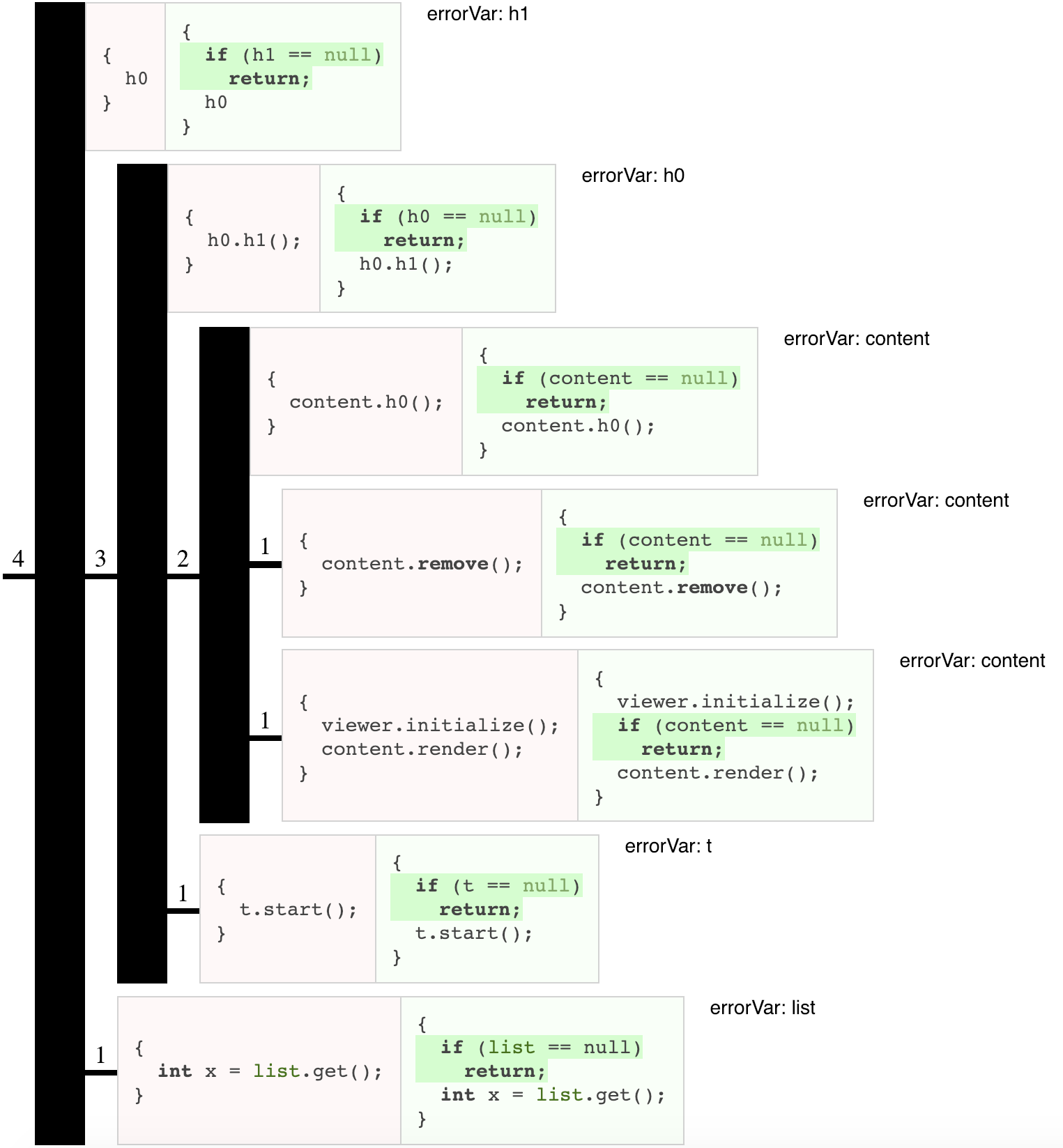}
	\caption{Dendrogram showing concrete edits merged into more abstract edit patterns. The vertical black bars correspond to levels in the hierarchy, and the edit pattern at the top of each black bar is the pattern obtained by anti-unification of the edit pattern connected by the smaller, thin black lines.}
	\label{fig:sample_dendro_full}
\end{figure*}

Before delving into our algorithm, consider the example hierarchy in Figure~\ref{fig:sample_dendro_full}, which visualizes a learned hierarchy of edit patterns as a so-called dendrogram.
Each pair of red and green boxes represents one edit pattern.
The dendrogram is derived from four concrete edits, which are shown as edit patterns without any holes in them.
The left-most edit pattern is the most general description of all four concrete edits, while all edit patterns in between represent different levels of generalization that comprise different subsets of the concrete edits.
The main benefit of keeping intermediate edit patterns, instead of greedily merging all edits into a single generalization, is that Getafix can choose the most suitable pattern to apply to a new piece of code (as explained in detail in Section~\ref{sec:patching}).

\subsubsection{Clustering Algorithm}

To obtain a hierarchy of edit patterns from a given set of concrete edits, Getafix performs a hierarchical, agglomerative clustering algorithm.
The algorithm maintains a working set $W$ of edit patterns.
Initially, this working set contains one edit pattern for each of the given concrete edits.
The main loop of the algorithm repeats the following steps until $W$ is reduced to a singleton set:
\begin{enumerate}
	\item Pick a pair $e_1, e_2$ of edit patterns from $W$.
	\item Generalize the edit patterns using anti-unification, which yields $e_3 = \antiunifyName(e_1,e_2)$.
	\item Remove $e_1$ and $e_2$ from $W$, and add $e_3$ to $W$ instead.
	\item Set $e_3$ as the parent of $e_1$ and $e_2$ in the hierarchy.
\end{enumerate}

\subsubsection{Merge Order}

An important component of our clustering algorithm is how to pick the pair $e_1, e_2$ of edit patterns to generalize.
Getafix aims at minimizing the loss of concrete information at each anti-unification step.
To this end, we exploit the fact that the anti-unification operation itself gives a partial order of edit patterns, which orders patterns by their level of abstraction.
Given this partial order of edit patterns, the algorithm picks a pair $e_1,e_2$ that produces a minimal generalization w.r.t. the partial order.
In the following, we first describe the partial order of edit patterns and then present an efficient approximation of it, which our Getafix implementation is based on.

The partial order of edit patterns is based on a partial order of tree patterns, which builds upon the anti-unification operation:

\begin{definition}[Partial order of tree patterns]\label{def:tprec}
	Anti-unification represents a least upper bound operation on \setTreeEx.
	The resulting semilattice $\langle \setTreeEx, \antiunifyName \rangle$ has \setTree as minimal elements and \hole{?}{k} (regardless of $k$) as its greatest element.
	$\langle \setTreeEx, \antiunifyName \rangle$ hence induces a partial order
	$\mpt$ on $\setTreeEx$ (read ``more precise than'') by defining $p \mpt p' \iff \antiunify{$p$}{$p'$} = p'$.
\end{definition}


The definition relies on equality of tree patterns.
We consider two tree patterns $p_1, p_2 \in \setTreeEx$ equal if they are structurally identical, modulo bijective substitution of hole indices.
For example,
$$\node{BinEx}{+}{\hole{Name}{0}, \hole{?}{1}} = \node{BinEx}{+}{\hole{Name}{5}, \hole{?}{3}}$$
$$\node{BinEx}{+}{\hole{Name}{0}, \hole{?}{1}} = \node{BinEx}{+}{\hole{Name}{1}, \hole{?}{0}}$$
$$\node{BinEx}{+}{\hole{Name}{0}, \hole{?}{1}} \neq \node{BinEx}{+}{\hole{Name}{2}, \hole{?}{2}}$$
	As a consequence, all single-hole patterns \hole{?}{k} are identical.

Finally, we define the partial order of edit patterns analogous to Definition \ref{def:tprec}:
\begin{definition}[Partial order of edit patterns]\label{def:eprec}~\\
	$\langle \setEditEx, \antiunifyName \rangle$ induces a partial order $\mpt$ on $\setEditEx$ (read ``more precise than'') by defining\\ $e \mpt e' \iff \antiunify{$e$}{$e'$} = e'$.
\end{definition}


\subsubsection{Approximations for Efficiency}
To ensure that Getafix scales to a large number of edits, our implementation performs three approximations.
First, since the partial order of edit patterns is computationally expensive, Getafix approximates it using the following steps, each acting as a tiebreaker for the previous step.
When picking edit pairs to merge, Getafix prefers generalizations that
\begin{enumerate}
	\item yield patterns without any unbound holes in the \after part over those with such holes,
	\item preserve more mappings between modified nodes,
	\item have fewer holes and holes generalizing away smaller subtrees,
	\item preserve more mappings between unmodified nodes with labels,
	\item preserve more error context (explained in Section~\ref{sec:context}), and
	\item preserve more mappings between unmodified nodes.
\end{enumerate}

Second, Getafix exploits the fact that after performing one merge, the new set of edit patterns and hence also the new set of merge candidates is largely unchanged.
To reduce the number of times that the same pair of potential merges is compared, the implementation uses the nearest-neighbor-chain algorithm~\cite{Benzecri1982NNChain}.

Third, because the clustering algorithm is quadratic in the number of concrete edits, we reduce the number of edits considered at once.
Specifically, the implementation partitions all given edits by the labels of their modified nodes in the \before and \after trees.
As edits for which the labels differ get least priority in the merge order, Getafix assigns them to separate partitions and constructs a hierarchy for each partition separately.
Finally, the approach generalizes the roots of these hierarchies to form a single dendrogram.

\subsection{Additional Context in Edit Patterns}
\label{sec:context}

\begin{figure}[t!]
	\centering
	\begin{subfigure}[b]{0.3\textwidth}
		\includegraphics[width=\textwidth]{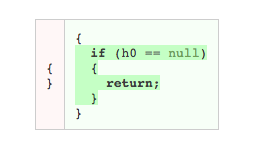}
		\caption{The hole \code{h0} is unbound.}
		\label{fig:unbound_edit_example}
	\end{subfigure}
	\hspace{0.02\textwidth}
	\begin{subfigure}[b]{0.3\textwidth}
		\includegraphics[width=\textwidth]{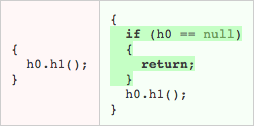}
		\caption{Code context binds \code{h0}.}
		\label{fig:unmodified_binding_context}
	\end{subfigure}
	\hspace{0.02\textwidth}
	\begin{subfigure}[b]{0.3\textwidth}
		\includegraphics[width=\textwidth]{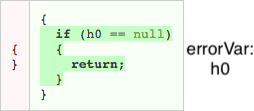}
		\caption{Error context binds \code{h0}.}
		\label{fig:error_var_binding_context}
	\end{subfigure}
	\caption{Edit pattern with an unbound hole and two ways of creating a bound version by adding context.}
	\label{fig:binding_an_unbound_edit}
\end{figure}

In addition to a bare description of modified code, the edit patterns learned by Getafix also include context information.
The motivation is twofold:
First, context makes edit patterns more specific, and hence, they match fewer code locations, which reduces the chance to produce invalid fixes.
Second, context information can bind otherwise unbound holes, which helps applying a pattern to new code.
Getafix uses two kinds of context: \emph{code context} and \emph{error context}.

\paragraph{Code context.}
Because Getafix starts with an entire spectrum of concrete edits, which include different amounts of code in addition to the changed AST nodes, the learned edit patterns also include contextual code.
Such contextual code helps deciding when a pattern is applicable, and it helps applying the pattern to new code.
Figure~\ref{fig:binding_an_unbound_edit} gives an example of an edit pattern with an unbound hole that becomes bound thanks to additional code context.
In edit pattern (a), \code{h0} is unbound in the \after part, leaving it unclear what to insert as \code{h0} when applying the pattern.
In contrast, edit pattern (b) binds \code{h0} by including unmodified code in the \before part.

\paragraph{Error context.}
The static analysis warning addressed by Getafix may also provide additional context information.
For example, the expression blamed for a \code{NullPointerException} gives a hint about how and where to apply a fix pattern.
Getafix optionally considers such hints in the form of an error variable.
While learning edit patterns, the approach then propagates this information, eventually obtaining patterns where a specific hole represents the error variable.
Figure~\ref{fig:binding_an_unbound_edit} gives an example for such error context.
Pattern (c) specifies that \code{h0} is the error variable.
As a result, Getafix can apply the pattern even though \code{h0} is not mentioned in the \before part.

\subsection{Comparison with Prior Work on Inferring Edit Patterns}
\label{sec:comparison greedy}

An alternative to our hierarchical clustering is a greedy clustering algorithm, as suggested for inferring recurring edit patterns~ \cite{Rolim2018LearningQF}.
That approach maintains a single representation of each cluster, and this representation does not contain any context information.
Even if one would add context information, the approach could add only context that is present in all of the edits in the training data, which is unlikely to learn edit patterns with the same level of context as Getafix.

For example, in Figure \ref{fig:sample_dendro_full}, the edit at the top of the hierarchy simply says to insert a check for null before some statement \code{h0}.
This is the only pattern that would be learned if we had only a single representation of this cluster.
Instead, hierarchical clustering enables Getafix to observe the pattern one level down, where the early return is inserted before a method call \code{h0.h1()}, which uses the variable that could be null.
Such extra context facilitates predicting a human-like fix.
Section~\ref{sec:eval ablation} compares our approach with greedy clustering.



\section{Applying and Ranking Fix Patterns}\label{sec:patching}

Given buggy source code and a set of learned fix patterns, the final step of Getafix produces a fixed version of the source code.
The main challenges of this step are (i) finding which patterns are applicable to a given piece of code (Section~\ref{sec:applying}) and (ii) deciding which of all fix candidates produced by applying these patterns to suggest to the developer (Section~\ref{sec:ranking}).

\subsection{Applying Edit Patterns to Buggy Code}
\label{sec:applying}

To find applicable edit patterns, Getafix matches tree patterns to the tree representation of buggy source code.

\begin{definition}[Matching]
	Let $p \in \setTreeEx$ and $t \in \setTree$. Then $p$ matches $t$ if there exists a function $\mathit{subst} : \setHole \funcarrow \setTree$ so that replacing every hole $h$ in $p$ with $\mathit{subst}(h)$ yields $t$. Intuitively, $\mathit{subst}$ captures the substitutions one has to perform to concretize $p$ into $t$.
\end{definition}

Given a specific edit pattern $p$ and a buggy tree $t$, the approach creates a fix using the following steps:
\begin{enumerate}
	\item Find a subtree $t_{sub}$ in $t$ that matches the \before part of $p$.
	\item Consistently instantiate all holes in the \before part and the \after part of $p$
	\item Replace the subtree $t_{sub}$ with the instantiated \after part.
\end{enumerate}

\begin{figure}[t]
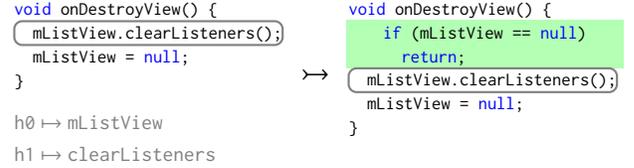

	\def\editColWidth{\dimexpr.2\textwidth}
	\hspace{0pt}\\*\noindent
	\begin{tabular}{@{}p{\editColWidth}@{}p{\dimexpr.05\textwidth}@{}p{\editColWidth}@{}}
		\begin{minipage}{\editColWidth}
			{\begin{bigcodeEdit}
void onDestroyView() {
  [[[mListView.clearListeners();]]]
  mListView = null;
}
			\end{bigcodeEdit}}
			{\footnotesize\color{gray}
				\texttt{h0} $\mapsto$ \texttt{mListView}\\
				\texttt{h1} $\mapsto$ \texttt{clearListeners}
			}
		\end{minipage}&
		\centering$\editarrow$&
		\begin{minipage}{\editColWidth}
			{\begin{bigcodeEdit}
void onDestroyView() {
  +++␣ if (mListView == null)   ␣+++
  +++␣   return;                ␣+++
  [[[mListView.clearListeners();]]]␣
  mListView = null;
}
			\end{bigcodeEdit}}
			\vspace{4pt}
		\end{minipage}
	\end{tabular}
	\caption{Example fix.}
	\label{fig:sample_fix}
\end{figure}

As an example, consider the code on the left of Figure~\ref{fig:sample_fix} and one of the edit patterns mentioned earlier:\footnote{To simplify the presentation, we will from now on represent edit patterns as pseudo-code instead of term notation.}
\begin{center}
	\edit{h0.h1();}{if (h0 == null) return; h0.h1();}
\end{center}
The \before part of the edit pattern matches the buggy code by substituting \code{h0} with \code{mListView} and \code{h1} with \code{clearListeners}.
Instantiating the \after part of the pattern with these substitutions and replacing the subtree that matches the \before part with the instantiated \after part results in the code on the right of Figure~\ref{fig:sample_fix}.

\subsection{Ranking Fix Candidates}
\label{sec:ranking}

By applying edit patterns, Getafix obtains a set $F_{cand}$ of applicable fixes, which we call fix candidates.
In practice, multiple patterns in a learned hierarchy of edit patterns may apply to a given piece of buggy code.
In fact, one of the main benefits of learning multiple edit patterns per bug category is to enable Getafix to select the most suitable pattern for a given situation.
However, having multiple applicable patterns leads to the challenge of deciding which fix to suggest to a developer.

To illustrate the problem, consider multiple learned fix patterns for null dereferences:

\vspace{.5em}
\begin{tabular}{rl}
	$p_1$: & \edit{\emph{(empty)}}{if (h0==null) \{ return; \}}\\
	&\hspace{0.5em} where \code{h0} is the error variable \\
	$p_2$: & \edit{h0.h1()}{h0!=null \&\& h0.h1()} \\
	$p_3$: & \edit{h0.h1();}{if (h0==null) \{ return; \} h0.h1();} \\
\end{tabular}
\vspace{.5em}

Some patterns match more often than others.
Patterns without unmodified nodes, such as $p_1$, match often and may even match multiple times within a given buggy piece of code.
For the code on the left of Figure~\ref{fig:sample_fix}, $p_1$ matches in unintended places, such as after \code{mListView.clearListeners()} or even after \code{mListView = null}.
In contrast, $p_2$ and $p_3$ are more specific, because they add the null check relative to existing code fragments.
Choosing which of these patterns is the most appropriate is a non-trivial problem.

We address the challenge of ranking fix candidates through a simple yet effective ranking algorithm that sorts fix candidates by their likelihood to be relevant, based on the human-written fixes that Getafix learns from.
Given a bug $b$ and a set $P$ of fix patterns, the problem is to rank all fix candidates $F_{cand}$ that result from applying some $p \in P$ to the code that contains $b$.
Let a tuple $(b, p, z)$ represent the application of pattern $p$ to the code that contains $b$ at source code location $z$.
The location $z$ is relative to the location where the static analyzer reports a warning about bug $b$.
For example, a location $z=-1$ means that the fix pattern is applied one line above the warning location, whereas $z=3$ means the fix pattern is applied three lines below the warning.
The goal of the ranking is that the probability $\bigProb{ \predHFix(b, p, z) \mid \predMatch(b, p, z) }$ is higher for fix candidates $(b, p, z)$ that appear higher up in the ranking, where \predMatch indicates that the pattern is applicable and \predHFix indicates that the fix is the same a human developer would implement.
Obviously, precisely computing \predHFix is impossible in general, because it depends on a human decision.
Instead, Getafix estimates the likelihood that a fix candidate will be accepted by a human based on the set $H$ of human fixes that the approach learns from.
To estimate how likely a fix candidate matches a human fix, Getafix computes three scores:
\begin{itemize}
	\item \emph{Prevalence score}.
	This score is higher for patterns that cover more concrete bug fixes in the set $H$ of human training fixes:
	$$s_{\mathit{prevalence}} = \dfrac{|\{ \mbox{instances of p in}~ H \}|}{|H|}$$
	The intuition is that a more common way of fixing a specific kind of bug is more likely to appeal to developers.

	\item \emph{Location score}.
	This score measures what fraction of bugs that were fixed with pattern $p$ were fixed by applying the pattern exactly $z$ lines away from the warning location:
	$$s_{\mathit{location}} = \dfrac{|\{ \mbox{instances of}~ p ~\mbox{at location}~ z ~\mbox{in}~ H \}|}{|\{ \mbox{instances of}~ p ~\mbox{in}~ H \}|}$$
	The motivation for this score is that some fix patterns typically occur only at specific locations.
	For example, fixing a null dereference by inserting an early return often occurs just above the location where the static analyzer warns about the potential null dereference.
	We describe below how Getafix computes a distribution of line distance for each learned pattern.

	\item \emph{Specialization score}.
	This score is higher for more specialized patterns, i.e., patterns that are applicable at fewer code locations.
$$s_{\mathit{specialized}} = \dfrac{|\{ \mbox{all AST nodes} \}|}{|\left\{ \mbox{\begin{tabular}{c}AST subtrees that match\\the \before part of $p$\end{tabular}} \right\}|}$$
	The intuition is that a more specific pattern contains more context information, and that if such a pattern matches, it is more likely to fit the given bug.

\end{itemize}
We estimate\\
$\bigProb{ \predHFix(b, p, z) \mid \predMatch(b, p, z) } \propto s_{\mathit{prevalence}} * s_{\mathit{location}} * s_{\mathit{specialized}}$, so the product of these scores is used for ranking fix candidates, with higher scores being ranked higher.
The location score is based on the distribution of locations, relative to the warning location, where a fix pattern is typically applied.
Getafix knows this location for each concrete fix in the training data, and then propagates the locations while generalizing fixes into fix patterns.
Specifically, we model the location as three statistical values tracked for each pattern during pattern learning.
First, we track the ratio of fixes applied before and fixes applied after the warning location.
This ratio is useful because some patterns typically apply either before or after the warning.
For concrete fixes, we initialize this ratio to 0, 1, and 0.5 if the human fix is above, below, or on the same line as the warning, respectively.
Second and third, we track two geometric distributions that model the line distance above and below the warning location.
For concrete fixes, we initialize these distributions as having their expected value exactly at the distance where the concrete fix is applied.
Whenever the clustering algorithm (Section~\ref{sec:clustering}) merges two edit patterns, it statistically propagates the three values from the two merged patterns to the resulting pattern.

As an example for the line distance distributions, reconsider fix pattern $p_1$ from above, which addresses a potential null deference by inserting an if statement that returns from the method if \code{h0} is null.
Because this pattern is typically applied before the warning location, the above/below ratio will be close to zero and the distribution for the line distance above will contain mostly small, positive integers.

\begin{example}[Ranking among candidates]~\\
	Assume that we are given the code from Figure \ref{fig:sample_fix} to fix and
	know the three fix patterns $p1$, $p2$ and $p3$ we motivated ranking with.
	Applying $p1$ results in three fix candidates $p1_1, p1_2, p1_3$ as mentioned earlier:
	The "early return" can be inserted above, between and below both statements.
	$p2$ and $p3$ only match on the buggy line, resulting in fix candidates $p2_1$ and $p3_1$.
	\begin{itemize}
		\item For $p1_1$ we assume the following scores:
			\begin{displaymath}
				s_{\mathit{prevalence}} = 0.03\quad
				s_{\mathit{location}} = 0.5\quad
				s_{\mathit{specialized}} = 20
			\end{displaymath}
			This would mean that we observed $p1$ being used 3\% of the time.
			Of those times, the statement was inserted right above the blamed line 50\% of the time.
			Furthermore the pattern matches one in twenty AST nodes.
		\item For $p1_2$ and $p1_3$ only $s_{\mathit{location}}$ changes since the other scores exclusively depend on the pattern used:
			\begin{displaymath}
				s_{\mathit{prevalence}} = 0.03\quad
				s_{\mathit{location}} = 0\quad
				s_{\mathit{specialized}} = 20
			\end{displaymath}
			Since insertion of an "early return" was never observed \textit{below} the buggy line, $s_{\mathit{location}}$ is zero.
		\item We assume $p2$ was observed 5\% of the time, almost always being applied to exactly the buggy line
			and $p2$ matches one in forty AST nodes on average.
			Resulting scores for $p2_1$:
			\begin{displaymath}
				s_{\mathit{prevalence}} = 0.05\quad
				s_{\mathit{location}} = 0.95\quad
				s_{\mathit{specialized}} = 40
			\end{displaymath}
		\item As $p3$ is more specialized than $p1$, we assume it was used for 2\% of fixes, but also matches only one in 200 AST nodes.
			The specialization contains the buggy method call, so it was also applied right there 90\% of the time.
			Resulting scores for $p3_1$:
			\begin{displaymath}
				s_{\mathit{prevalence}} = 0.02\quad
				s_{\mathit{location}} = 0.9\quad
				s_{\mathit{specialized}} = 200
			\end{displaymath}
	\end{itemize}
	We conclude that Getafix would rank candidates in order $p3_1$, $p2_1$, $p1_1$ and so on due to respective overall scores $3.6$, $1.9$, $0.3$ and $0$.
\end{example}

\subsection{Comparison with Other Ranking Techniques}
\label{sec:comparison ranking}

The ranking step of Getafix relates to prior work on ranking fix candidates in the context of automated program repair.
In principle, other ways of ranking may be plugged into Getafix, either to replace and to complement our current approach.
One candidate for such an enhancement is the Prophet ranking~\cite{Long2016Automatic}, which uses a set of over 3000 features of code and code changes to predict which fixes are more likely to be adequate.
We did not add Prophet to our current prototype because the existing Getafix ranking works well in practice (Section~\ref{sec:eval ranking}) and because adapting Prophet's over 3000 features to the Java programming language is non-trivial.
Conceptually, there are two advantages of our ranking over Prophet.
First, Getafix ranks candidate fixes based on human fixes for the same kind of bug, whereas Prophet relies on a generic model of how ``natural'' a code change is.
Learning to rank from fixes for the same bug category allows Getafix to specialize to bug category-specific ``features''.
Second, our ranking is defined based on ASTs and line numbers, i.e., in a language-agnostic way, making it easy to apply the approach to another language.

\section{Evaluation}\label{sec:eval}

We address the following research questions:
\begin{itemize}
	\item \textbf{RQ1}: How effective is Getafix at predicting human-like fixes? (Section~\ref{sec:eval fixes})

	\item \textbf{RQ2}: How effective is the ranking of fix candidates? (Section~\ref{sec:eval ranking})

	\item \textbf{RQ3}: How does Getafix compare to simpler variants of the approach? (Section~\ref{sec:eval ablation})

	\item \textbf{RQ4}: How does the amount of available training data affect the approach? (Section~\ref{sec:eval data})

	\item \textbf{RQ5}: How efficient is the approach? (Section~\ref{sec:eval efficiency})

	\item \textbf{RQ6}: How effective is Getafix in an industrial software development setting? (Section~\ref{sec:eval deployment})

\end{itemize}

\subsection{Experimental Setup}
To address RQ1 to RQ5, we evaluate Getafix with pairs of \before and \after Java files that address six kinds of problems.
Table~\ref{tab:accuracy} lists the bug categories and the static analysis tool that detects instances of these categories.
We use bugs detected by Infer\doubleblind{}{\footnote{\urlx{https://code.fb.com/developer-tools/open-sourcing-facebook-infer-identify-bugs-before-you-ship/}}}~\cite{calcagno2015moving} and Error Prone~\cite{Aftandilian2012}, two state-of-the-art static analyzers for Java.
To gather fixes of potential NullPointerExceptions, we gather fixes made by developers at \doubleblind{CompanyX}{Facebook} as a reaction to warnings reported by Infer.
For the other five bug categories, we gather fixes made by developers of open-source developers by searching for commits mentioning those warnings.
Since some commits also contain code changes unrelated to fixing the specific bug category, we semi-automatically split and filter commits into pairs of \before and \after files, so that each pair fixes exactly one instance of the bug category.
Overall, the data set contains 1,268 pairs of files.

Unless otherwise mentioned, all experiments are performed in a 10-fold cross-validation setup.
We configure the pattern learning step to drop fix patterns representing less than 1\% of the training set, to avoid polluting the prediction with very uncommon patterns.


\subsection{Effectiveness in Finding Human-like Fixes}
\label{sec:eval fixes}

We measure Getafix's effectiveness in finding human-like fixes by checking whether a predicted fix exactly matches the known human fix.
This comparison neglects empty lines but is otherwise precise, so even the addition of a comment or different whitespace choices are judged as a mismatch.
The rationale is that we want to make sure the suggestions are acceptable to developers.
This measure of success underapproximates the effectiveness of Getafix, because there may be other fix suggestions that a developer would have accepted than exactly the fix that the developer has applied, e.g., a semantically equivalent fix.
We measure both top-1 accuracy, i.e., the percentage of fixes that Getafix predicts as the top-most suggestion, top-5 accuracy, i.e., the percentage of fixes that Getafix predicts as one of the first five suggestions.

\begin{table*}
	\caption{Accuracy of predicting exactly the human fix for different kinds of bugs.}
	\label{tab:accuracy}
	\small
\begin{tabular}{llrp{.5em}rlp{.5em}rl}
\toprule
Bug category & Static analyzer & Examples & \multicolumn{3}{c}{Top-1 accuracy} & \multicolumn{3}{c}{Top-5 accuracy} \\
\midrule
NullPointerException & Infer & 804 && 94 & (12\%) && 156 & (19\%)\\
BoxedPrimitiveConstructor & Error Prone & 260 && 236 & (91\%) && 238 & (92\%)\\
ClassNewInstance & Error Prone & 21 && 6 & (29\%) && 11 & (52\%)\\
DefaultCharSet & Error Prone & 55 && 28 & (51\%) && 42 & (76\%)\\
OperatorPrecedence & Error Prone & 49 && 6 & (12\%) && 30 & (61\%)\\
ReferenceEquality & Error Prone & 79 && 22 & (28\%) && 53 & (67\%)\\
\midrule
Total & & 1,268 && 381 &&& 526 \\
\bottomrule
\end{tabular}
\end{table*}

Table~\ref{tab:accuracy} shows the accuracy results for all six bug categories.
Depending on the bug category, Getafix suggests the correct fix as the top-most suggestion for 12\% to 91\% of all fixes, fixing 381 out of a total of 1,268 bugs.
The top-5 accuracy is even higher, ranging between 19\% and 92\%, fixing a total of 526 out of 1,268 bugs.

The differences between bug categories reflect how diverse and complex typical fixes for the different kinds of static analysis warnings are.
For the bug category with highest accuracy, BoxedPrimitiveConstructor, there exists a relatively small set of recurrent fix patterns, which Getafix easily learns and applies.
In contrast, the problem of automatically fixing potential NullPointerExceptions is harder, as developers use a wide range of fix patterns.
For example, there are at least two fundamentally different ways to address a null deference warning:
Either the developer recognizes that the blamed expression may indeed be null at runtime, or the developer is certain that, for reasons inaccessible to the static analyzer, the expression cannot be null.
Depending on this decision, the fix will either change logic or control flow to behave reasonably in case of a null value, as, e.g., shown in Figure~\ref{fig:NPEexamples} or add a non-null assertion that convinces the static analyzer about non-nullability.
Both cases provide a range of different fix patterns, which Getafix learns and applies successfully in some, but not all cases.
Another factor influencing the accuracy of Getafix is, as for any learning-based approach, the amount of available training data, which we discuss separately in Section~\ref{sec:eval data}.

\subsubsection{Examples of Correctly Predicted Fixes}

To illustrate the strengths and limitations of Getafix, let us consider some examples.
Figure~\ref{fig:NPEexamples} from the introduction gives three examples of fixes for NullPointerExceptions that Getafix predicts as the top-most suggestion.
These examples illustrate that the approach not only predicts a wide range of different strategies for fixing bugs, but also selects the most suitable strategy for the given code.
For example, the first fix in Figure~\ref{fig:NPEexamples}, which adds a check to an existing conditional, makes sense only when the code surrounding the bug already has a conditional.
The hierarchy of fix patterns, many of which provide some code context, allows our ranking to prioritize the most suitable patterns.

\begin{figure*}[t]
	\begin{subfigure}[b]{\textwidth}
	\def\editColWidth{\dimexpr.3\textwidth}
	\hspace{0pt}\\*\noindent
  \centering
\begin{tabular}{@{}p{\editColWidth}@{}p{\dimexpr.1\textwidth}@{}p{\editColWidth}@{}}
	\begin{minipage}{\editColWidth}
		{\begin{bigcodeEdit}
@Override
public String toString() {
  return ---new Integer(key)---.toString();
}
		\end{bigcodeEdit}}
	\end{minipage}&
	\centering$\editarrow$&
	\begin{minipage}{\editColWidth}
		{\begin{bigcodeEdit}
@Override
public String toString() {
  return +++Integer.valueOf(key)+++.toString();
}
\end{bigcodeEdit}}
	\end{minipage}
\end{tabular}
\caption{Fix for warning about boxing a primitive value into a fresh object.}
\label{fig:BoxedPrimitiveConstructor}
	\end{subfigure}

\vspace{1em}

	\begin{subfigure}[b]{\textwidth}
	\def\editColWidth{\dimexpr.3\textwidth}
	\hspace{0pt}\\*\noindent
  \centering
	\begin{tabular}{@{}p{\editColWidth}@{}p{\dimexpr.1\textwidth}@{}p{\editColWidth}@{}}
		\begin{minipage}{\editColWidth}
			{\begin{bigcodeEdit}
// Ensure that the text hasn't changed.
assert ---view.getText().toString()---
    ---== metadata.getString()---
    : "The text '"
    + view.getText().toString()
    + "' has been modified"
			\end{bigcodeEdit}}
		\end{minipage}&
		\centering$\editarrow$&
		\begin{minipage}{\editColWidth}
			{\begin{bigcodeEdit}
// Ensure that the text hasn't changed.
assert +++view.getText().toString().equals(metadata.getString())+++

    : "The text '"
    + view.getText().toString()
    + "' has been modified"
			\end{bigcodeEdit}}
		\end{minipage}
	\end{tabular}
\caption{Fix for warning about comparison with reference equality.}
\label{fig:ReferenceEquality}
\end{subfigure}

	\caption{Correctly predicted fixes for BoxedPrimitiveConstructor and ReferenceEquality warnings.}
	\label{fig:fixExamples}
\end{figure*}

Figure~\ref{fig:fixExamples} shows correctly predicted fixes for two other kinds of bug categories.
The fix in Figure~\ref{fig:BoxedPrimitiveConstructor} addresses a warning about boxing a primitive value into a fresh object, which is inefficient compared to the fixed code that implicitly uses a cached instance of frequently used values.
The fix in Figure~\ref{fig:ReferenceEquality} replaces an expression that compares two objects via reference equality with a call to \code{equals}.
Getafix learns these fix patterns because fixes for these bug categories tend to be repetitive.
Yet, finding an appropriate fix is non-trivial because there are different fix variants.
For example, other fixes for the BoxedPrimitiveConstructor warning address different combinations of wrapped types, e.g., floats or strings, and different types of object wrappers, e.g., \code{Float} or \code{Boolean}.
Likewise, other fixes for the ReferenceEquality warning use an inequality check instead of an equality check, they might use \code{Objects.equals()} to do the equality check, or they may be part of a more complex expression.

\subsubsection{Examples of Missed Fixes}

\begin{figure*}[t]
	\begin{subfigure}[b]{\textwidth}
		\def\editColWidth{\dimexpr.3\textwidth}
		\hspace{0pt}\\*\noindent
    \centering
		\begin{tabular}{@{}p{\editColWidth}@{}p{\dimexpr.1\textwidth}@{}p{\editColWidth}@{}}
			\begin{minipage}{\editColWidth}
				{\begin{bigcodeEdit}
// ...

bar = x.foo();
// ...
				\end{bigcodeEdit}}
			\end{minipage}&
			\centering$\editarrow$&
			\begin{minipage}{\editColWidth}
				{\begin{bigcodeEdit}
// ...
+++if (x == null) throw new IllegalStateException("custom message");+++
bar = x.foo();
// ...
				\end{bigcodeEdit}}
			\end{minipage}
		\end{tabular}
		\caption{Fix for potential null deference.}
		\label{fig:NPEmissed}
	\end{subfigure}

	\vspace{.5em}

	\begin{subfigure}[b]{\textwidth}
		\def\editColWidth{\dimexpr.3\textwidth}
    \hspace{0pt}\\*\noindent
    \centering
		\begin{tabular}{@{}p{\editColWidth}@{}p{\dimexpr.1\textwidth}@{}p{\editColWidth}@{}}
			\begin{minipage}{\editColWidth}
				{\begin{bigcodeEdit}

// ...
return clazz.cast(---in.newInstance()---);
// ...


// ...
				\end{bigcodeEdit}}
			\end{minipage}&
			\centering$\editarrow$&
			\begin{minipage}{\editColWidth}
				{\begin{bigcodeEdit}
+++import java.lang.reflect.InvocationTargetException;+++
// ...
  return clazz.cast(+++in.getConstructor().newInstance()+++);
// ...
+++} catch (InvocationTargetException e) {+++
+++} catch (NoSuchMethodException e) {+++
// ...
				\end{bigcodeEdit}}
			\end{minipage}
		\end{tabular}
		\caption{Fix for ClassNewInstance warning.}
		\label{fig:ClassNewInstancemissed}
	\end{subfigure}

	\caption{Fixes that Getafix could not predict correctly.}
	\label{fig:nofixExamples}
\end{figure*}

To better understand the limits of Getafix, Figure~\ref{fig:nofixExamples} shows two fixes that the approach does not find.
Figure~\ref{fig:NPEmissed} fixes a potential null dereference by throwing an IllegalStateException in case a variable is null.
While finding this additional statement is well in scope Getafix, it fails to predict the custom error message that is passed as a string to the IllegalStateException.
Since the message is application-specific, learning to predict it is a hard problem.
Figure~\ref{fig:ClassNewInstancemissed} fixes a warning about the \code{newInstance} API, which is discouraged because it bypasses exception checking.
Fixes of this problem often have to adapt the exception handling around the actual fix location.
Since these adaptations heavily depend on the existing exception handling, Getafix often fails to predict exactly the correct fix.
This limitation could likely be mitigated by a larger set of training examples, which would enable Getafix to learn popular exception handling scenarios.
Other reasons why Getafix misses fixes include fixes that rename variables, simply remove buggy code, or that address a deeper root cause of the problem.


\subsection{Effectiveness of Ranking}
\label{sec:eval ranking}

\begin{figure*}[t]
  \begin{subfigure}[b]{0.32\textwidth}
    \begin{tikzpicture}[thick,scale=0.6]
      \begin{axis}[
        cycle list name=color,
        symbolic x coords={1,2,3,4,5,6,7,8,$\infty$},
        xtick={1,2,3,4,5,6,7,8,$\infty$},
        yticklabel=\pgfmathprintnumber{\tick}\%,
        enlargelimits=true,
        xmin=1, xmax=$\infty$,
        ymajorgrids,
        width=3.2in,
        height=2.5in,
      ]
      \draw[black] decorate [decoration={zigzag}] {([xshift=6pt] axis cs:8,110) -- ([xshift=6pt] axis cs:8,-10)};
      \addplot coordinates {(1,11.59) (2,14.15) (3,16.23) (4,17.46) (5,18.92) (6,19.88) (7,20.76) (8,21.57)};
      \node [right] at (axis cs: 8, 21.57) {\hspace{4pt} 29.7\%};
      \addplot coordinates {(1,4.23) (2,4.54) (3,5.72) (4,6.84) (5,8.02) (6,8.71) (7,9.95) (8,10.82)};
      \node [right] at (axis cs: 8, 10.82) {\hspace{4pt} 40.7\%};
      \addplot coordinates {(1,9.92) (2,11.60) (3,12.13) (4,12.62) (5,13.09) (6,13.25) (7,13.65) (8,13.71)};
      \node [right] at (axis cs: 8, 13.71) {\hspace{4pt} 14.1\%};
      \end{axis}
    \end{tikzpicture}
    \caption{NullPointerException}
  \end{subfigure}
  \begin{subfigure}[b]{0.32\textwidth}
    \begin{tikzpicture}[thick,scale=0.6]
      \begin{axis}[
        cycle list name=color,
        symbolic x coords={1,2,3,4,5,6,7,8,$\infty$},
        xtick={1,2,3,4,5,6,7,8,$\infty$},
        yticklabel=\pgfmathprintnumber{\tick}\%,
        enlargelimits=true,
        xmin=1, xmax=$\infty$,
        ymajorgrids,
        width=3.2in,
        height=2.5in,
      ]
      \draw[black] decorate [decoration={zigzag}] {([xshift=6pt] axis cs:8,110) -- ([xshift=6pt] axis cs:8,-10)};
      \addplot coordinates {(1,90.64) (2,91.15) (3,91.67) (4,91.67) (5,91.67) (6,91.67) (7,91.67) (8,91.67)};
      \node [right] at (axis cs: 8, 91.67) {\hspace{4pt} 91.7\%};
      \addplot coordinates {(1,90.77) (2,91.15) (3,91.41) (4,92.05) (5,92.05) (6,92.05) (7,92.05) (8,92.05)};
      \node [right] at (axis cs: 8, 92.05) {\hspace{4pt} 92.1\%};
      \addplot coordinates {(1,91.15) (2,91.15) (3,91.15) (4,91.15) (5,91.15) (6,91.15) (7,91.15) (8,91.15)};
      \node [right] at (axis cs: 8, 91.15) {\hspace{4pt} 91.2\%};
      \end{axis}
    \end{tikzpicture}
    \caption{BoxedPrimitiveConstructor}
  \end{subfigure}
   \begin{subfigure}[b]{0.32\textwidth}
    \begin{tikzpicture}[thick,scale=0.6]
      \begin{axis}[
        cycle list name=color,
        symbolic x coords={1,2,3,4,5,6,7,8,$\infty$},
        xtick={1,2,3,4,5,6,7,8,$\infty$},
        yticklabel=\pgfmathprintnumber{\tick}\%,
        enlargelimits=true,
        xmin=1, xmax=$\infty$,
        ymajorgrids,
        width=3.2in,
        height=2.5in,
      ]
      \draw[black] decorate [decoration={zigzag}] {([xshift=6pt] axis cs:8,110) -- ([xshift=6pt] axis cs:8,-10)};
      \addplot coordinates {(1,28.57) (2,52.38) (3,52.38) (4,52.38) (5,52.38) (6,52.38) (7,52.38) (8,52.38)};
      \node [right] at (axis cs: 8, 52.38) {\hspace{4pt} 52.4\%};
      \addplot coordinates {(1,14.29) (2,14.29) (3,14.29) (4,14.29) (5,14.29) (6,14.29) (7,14.29) (8,14.29)};
      \node [right] at (axis cs: 8, 14.29) {\hspace{4pt} 52.4\%};
      \addplot coordinates {(1,23.81) (2,23.81) (3,30.95) (4,38.10) (5,38.10) (6,38.10) (7,38.10) (8,38.10)};
      \node [right] at (axis cs: 8, 38.10) {\hspace{4pt} 38.1\%};
      \end{axis}
    \end{tikzpicture}
    \caption{ClassNewInstance}
  \end{subfigure}
    \begin{subfigure}[b]{0.32\textwidth}
    \begin{tikzpicture}[thick,scale=0.6]
      \begin{axis}[
        cycle list name=color,
        symbolic x coords={1,2,3,4,5,6,7,8,$\infty$},
        xtick={1,2,3,4,5,6,7,8,$\infty$},
        yticklabel=\pgfmathprintnumber{\tick}\%,
        enlargelimits=true,
        xmin=1, xmax=$\infty$,
        ymajorgrids,
        width=3.2in,
        height=2.5in,
      ]
      \draw[black] decorate [decoration={zigzag}] {([xshift=6pt] axis cs:8,110) -- ([xshift=6pt] axis cs:8,-10)};
      \addplot coordinates {(1,52.73) (2,63.64) (3,70.91) (4,75.45) (5,75.45) (6,75.45) (7,76.36) (8,77.27)};
      \node [right] at (axis cs: 8, 77.27) {\hspace{4pt} 85.5\%};
      \addplot coordinates {(1,0.00) (2,0.91) (3,0.91) (4,0.91) (5,0.91) (6,1.82) (7,1.82) (8,1.82)};
      \addplot coordinates {(1,0.00) (2,0.00) (3,0.00) (4,1.82) (5,1.82) (6,1.82) (7,1.82) (8,1.82)};
      \node [right] at (axis cs: 8, 1.82) {\hspace{4pt} 1.8\%};
      \end{axis}
    \end{tikzpicture}
    \caption{DefaultCharSet}
  \end{subfigure}
  \begin{subfigure}[b]{0.32\textwidth}
    \begin{tikzpicture}[thick,scale=0.6]
      \begin{axis}[
        cycle list name=color,
        symbolic x coords={1,2,3,4,5,6,7,8,$\infty$},
        xtick={1,2,3,4,5,6,7,8,$\infty$},
        yticklabel=\pgfmathprintnumber{\tick}\%,
        enlargelimits=true,
        xmin=1, xmax=$\infty$,
        ymajorgrids,
        width=3.2in,
        height=2.5in,
      ]
      \draw[black] decorate [decoration={zigzag}] {([xshift=6pt] axis cs:8,110) -- ([xshift=6pt] axis cs:8,-10)};
      \addplot coordinates {(1,13.27) (2,32.65) (3,46.43) (4,53.57) (5,58.67) (6,58.67) (7,58.67) (8,58.67)};
      \node [right] at (axis cs: 8, 58.67) {\hspace{4pt} 63.8\%};
      \addplot coordinates {(1,14.29) (2,30.61) (3,42.86) (4,44.90) (5,46.94) (6,46.94) (7,46.94) (8,50.00)};
      \node [right] at (axis cs: 8, 50.00) {\hspace{4pt} 61.2\%};
      \addplot coordinates {(1,14.29) (2,14.29) (3,14.29) (4,14.29) (5,14.29) (6,14.29) (7,14.29) (8,14.29)};
      \node [right] at (axis cs: 8, 14.29) {\hspace{4pt} 14.3\%};
      \end{axis}
    \end{tikzpicture}
    \caption{OperatorPrecedence}
  \end{subfigure}
  \begin{subfigure}[b]{0.32\textwidth}
    \begin{tikzpicture}[thick,scale=0.6]
      \begin{axis}[
        cycle list name=color,
        symbolic x coords={1,2,3,4,5,6,7,8,$\infty$},
        xtick={1,2,3,4,5,6,7,8,$\infty$},
        yticklabel=\pgfmathprintnumber{\tick}\%,
        enlargelimits=true,
        xmin=1, xmax=$\infty$,
        ymajorgrids,
        width=3.2in,
        height=2.5in,
      ]
      \draw[black] decorate [decoration={zigzag}] {([xshift=6pt] axis cs:8,110) -- ([xshift=6pt] axis cs:8,-10)};
      \addplot coordinates {(1,28.48) (2,37.34) (3,46.20) (4,54.75) (5,67.09) (6,69.30) (7,69.62) (8,69.62)};
      \node [right] at (axis cs: 8, 69.62) {\hspace{4pt} 74.4\%};
      \addplot coordinates {(1,29.11) (2,31.65) (3,41.77) (4,45.57) (5,48.10) (6,49.37) (7,49.37) (8,49.37)};
      \node [right] at (axis cs: 8, 49.37) {\hspace{4pt} 50.6\%};
      \addplot coordinates {(1,26.58) (2,35.44) (3,37.97) (4,37.97) (5,37.97) (6,37.97) (7,37.97) (8,37.97)};
      \node [right] at (axis cs: 8, 37.97) {\hspace{4pt} 38.0\%};
      \end{axis}
    \end{tikzpicture}
    \caption{ReferenceEquality}
  \end{subfigure}
  \includegraphics[width=.3\linewidth]{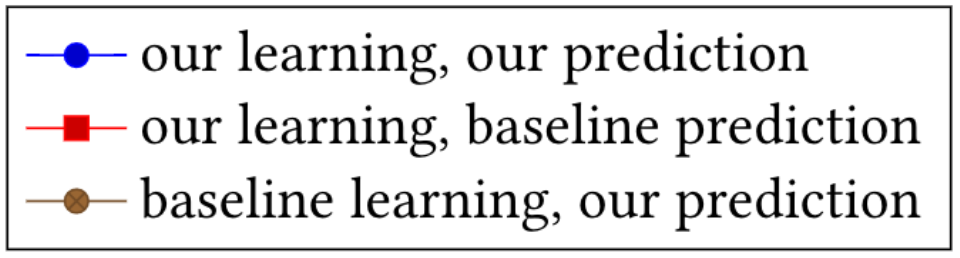}
  \caption{Fraction of human fixes (y axis) covered by top-$k$ (x axis) highest ranked fix candidates.}
  \label{fig:ablation}
\end{figure*}

An important component of Getafix is the ranking of candidate fixes, which enables the approach to decide on a small number of fix suggestions, without passing all candidates to any computationally expensive validation step.
Figure~\ref{fig:ablation} evaluates the effectiveness of the ranking, showing as a blue line (``our learning, our prediction'') how many of all fixes Getafix predicts correctly within the $k$ highest ranked suggestions per bug.
The percentage shown on the right of each plot, with $k=\infty$, indicates for how many bugs the correct fix is somewhere in the entire set of fix candidates.
The results show that Getafix's ranking effectively pushes the correct fix toward the top of all candidate fixes, covering a large fraction of all fixes it can find within the first few suggestions.


\subsection{Ablation Study}
\label{sec:eval ablation}

To evaluate the effectiveness of both our hierarchical learning algorithm and our ranking technique,
we compare them against baseline approaches.
As a baseline for learning, we use the greedy clustering algorithm used by Revisar~\cite{Rolim2018LearningQF}.
Revisar first partitions concrete edits by their "$d$-cap", which means that all concrete edits in one partition will have ASTs that are identical for all nodes up to depth $d-1$.
Within each partition, Revisar then considers one edit after another, attempting to cluster them into a pattern that becomes more and more general.
When the addition of an edit to a cluster would result in a pattern that contains an unbound hole, this edit instead forms a new cluster.
We conservatively approximate this behavior by selecting patterns from our hierarchy of patterns that could have been found by greedy clustering if concrete edits were iterated in an ideal order.
We focus our comparison on $1$-cap, as it results in more general patterns than $d > 1$.
We also drop additional context (see Section \ref{sec:comparison greedy}).
As a baseline for prediction, we learn a hierarchy of fix patterns as described in Section~\ref{sec:clustering}, but then always pick the most common pattern of the hierarchy, instead of applying our ranking from Section~\ref{sec:ranking}.
In case this pattern matches at multiple code locations, we apply it to the location closest to the reported issue.
When evaluating the top-k accuracy of this baseline, we repeatedly pick the next most common pattern.

Figure~\ref{fig:ablation} compares Getafix to the two baselines.
For five out of six bug categories, both hierarchical learning and our multi-score ranking are clearly worthwhile.
For BoxedPrimitiveConstructor, all three approaches perform roughly the same, which we attribute to the fact that this category has relatively few ways of fixing the bug.
The most drastic drop in accuracy can be observed without the hierarchical clustering, and hence without the valuable inner nodes of the dendrogram.
The DefaultCharSet category sees particularly large drop in accuracy because most human fixes involve adding an import statement but the baseline learning and prediction do not support learning and applying multiple learned patterns together.


\subsection{Influence of Available Training Data}
\label{sec:eval data}

\begin{figure}[t]
  \begin{tikzpicture}[scale=0.8]
    \begin{axis}[
    cycle list name=color,
    legend style={at={(1.77, 0.5)},anchor=east},
    legend cell align={left},
    symbolic x coords={0.1,0.2,0.3,0.4,0.5,0.6,0.7,0.8,0.9},
    xtick={0.1,0.2,0.3,0.4,0.5,0.6,0.7,0.8,0.9},
    yticklabel=\pgfmathprintnumber{\tick}\,\%,
    enlargelimits=true,
    xmin=0.1, xmax=0.9,
    ymajorgrids,
    xlabel=$t$,
    ylabel=Accuracy,
    width=2.5in,
    height=2.2in,
    ]
    \addlegendentry{Top-$1$ accuracy}
    \addplot coordinates {			(0.1,8.84) 	(0.2,8.77) 	(0.3,7.81) 	(0.4,9.65) 	(0.5,9.16) 	(0.6,10.53) 	(0.7,10.24) 	(0.8,10.85) 	(0.9,11.62) 	};
    \addlegendentry{Top-$5$ accuracy}
    \addplot coordinates {			(0.1,14.11) 	(0.2,16.77) 	(0.3,16.28) 	(0.4,18.70) 	(0.5,18.02) 	(0.6,18.96) 	(0.7,18.55) 	(0.8,18.78) 	(0.9,18.89) 	};
    \addlegendentry{Top-$\infty$ accuracy}
    \addplot coordinates {			(0.1,18.53) 	(0.2,24.68) 	(0.3,25.17) 	(0.4,27.93) 	(0.5,28.28) 	(0.6,28.89) 	(0.7,28.86) 	(0.8,29.61) 	(0.9,29.65) 	};
    \end{axis}
  \end{tikzpicture}
  \caption{Accuracy of top-$1$, top-$5$ and top-$\infty$ predictions (fraction of predicted human NullPointerException fixes) depending on fraction $t$ of data used for training.}
  \label{fig:tdatafraction}
\end{figure}
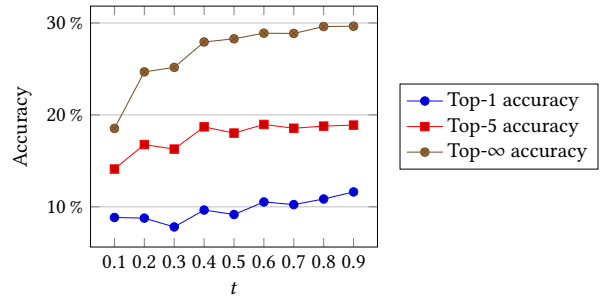

To measure the influence of the amount of available training data, we run Getafix with subsets of the 804 human fixes for null dereference bugs.
Figure \ref{fig:tdatafraction} shows the accuracy of reproducing the human fix among top-k fix suggestions, depending on the fraction $t$ of data used for training.
We use for validation the samples not used for training, and repeat the experiment for each $t$ such that each fix was used at least eight times for both training and validation.
As expected for a learning-based approach, more training data yields higher accuracy.
The reason is that Getafix learns more patterns, as well as more accurate and representative metadata for those patterns (frequency, line distance information, etc.), which improves the ranking.


\subsection{Efficiency}
\label{sec:eval efficiency}

\begin{table*}
\caption{Time taken by Getafix for 10-fold experiment of training and prediction.}
\label{tab:efficiency}
\small
\begin{tabular}{lrrrr} \toprule
  Bug category & Examples & Training time & Prediction time & Single prediction\\ \midrule
  NullPointerException & 804 & 762s & 3132s & 3.9s \\
  BoxedPrimitiveConstructor & 260 & 164s & 442s & 1.7s \\
  ClassNewInstance & 21 & 60s & 195s & 9.2s \\
  DefaultCharSet & 55 & 76s & 240s & 4.4s \\
  OperatorPrecedence & 49 & 70s & 138s & 2.8s \\
  ReferenceEquality & 79 & 114s & 563s & 7.2s \\
  \bottomrule
\end{tabular}
\end{table*}

Table~\ref{tab:efficiency} lists the total time spent on training and prediction for each 10-fold experiment.
A more realistic setup would correspond to one fold of this experiment.
Breaking down the total time to individual predictions of fixes (last column), we find that Getafix takes between 1.7 seconds and 9.2 seconds on average to predict a fix.
These experiments where conducted on a single machine with 114GB of RAM and Intel Skylake processors with 24 cores running at 2.4GHz.

\subsection{Real-World Deployment: Auto-Fix Suggestions for Infer Bug Reports}
\label{sec:eval deployment}

%


Getafix is deployed in \doubleblind{CompanyX}{Facebook} to automatically suggest fixes for null dereference bugs reported by Infer.
For this deployment, 
we dropped all patterns that add assertions to the code or that just remove code,
because these patterns produce a plethora of candidates that are hard to rank.
Before displaying auto-fix suggestions, we rerun Infer to ensure that previously reported warnings have disappeared.
Getafix performs this validation only for the top-ranked predicted fix, which is the only
one shown to the developer (if the validation passes); showing additional fix candidates would require
additional runs of Infer, and besides, may be distracting for the developer.
For 84\% of these validation runs, the top-most predicted fix removes the previously reported Infer warning, and Getafix hence suggests the fix to the developers.

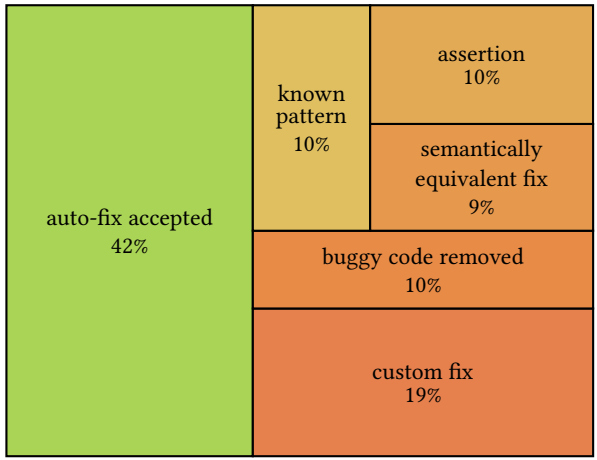
\begin{figure}[t]
	\small
  \begin{tikzpicture}[xscale=1.3,yscale=1]
    \pie[
      square,
      text=inside,
      color={%
        {rgb:red,1;green,2;yellow,1;white,2},%
        {rgb:red,5;green,1;yellow,1;white,3},%
        {rgb:red,5;green,1;yellow,2;white,3},%
        {rgb:red,2;green,1;yellow,2;white,3},%
        {rgb:red,4;green,1;yellow,2;white,3},%
        {rgb:red,3;green,1;yellow,2;white,3}%
      }
    ]
    {
      42/auto-fix accepted,
      19/custom fix,
      10/buggy code removed,
      10/known\\pattern,
      9/semantically\\equivalent fix,
      10/assertion
    }
  \end{tikzpicture}
  \caption{Reactions to about 250 null deference warnings for which Getafix suggested a fix.}
  \label{fig:refractions}
\end{figure}

Within three months of deployment, developers addressed around 250 null dereference warnings that we showed a fix suggestion for.
Figure~\ref{fig:refractions} categorizes the reactions of developers:
The developers directly accepted 106 suggested fixes (42\%).
These fixes include many non-trivial code changes (e.g., see Figure~\ref{fig:NPEexamples}), i.e., Getafix helped save precious developer time.
Remarkably, the acceptance rate of 42\% is significantly higher than the top-1 accuracy during the experiments reported in Section~\ref{sec:eval fixes}.
This difference is because developers are likely to accept more than one possible fix for a given bug, whereas Section~\ref{sec:eval fixes} compares the predicted fix against a single known human fix.

For those fix suggestions that were not immediately accepted, Figure~\ref{fig:refractions} categorizes them by their potential to be claimed by Getafix.
In 10\% of the cases, the developers eventually wrote a fix that Getafix knows, but that it did not rank highest and hence did not suggest.
Better ranking could address these cases.
Another 10\% of the time, the developer added an assertion that prevents the static analysis warning.
As indicated earlier, we have intentionally not rolled out such patterns so far, to not encourage developers to suppress analysis warnings.
Around 9\% of the time, a semantically equivalent fix to that suggested by Getafix was written.
For example, for a bug in an \code{else} block, the developer turned the block it into an \code{else if}, rather than creating a new \code{if}.
Learning more specialized patterns from more training data, or manual interpolation of learned patterns, will help to address such cases.
In another 10\% of the cases, the issue was resolved by just removing the code that contains the warning.
Finally, the remaining human fixes were mostly custom and probably cannot be expressed as a recurring fix pattern.


In addition to warnings about null dereferences, we have deployed Getafix for two other Infer warnings, Field Not Nullable and Return Not Nullable.
The acceptance rate of fixes suggested by Getafix is around 60\% for both warnings.
Moreover, displaying an auto-fix next to a warning resulted in a 4\% and 12\% higher fix rate, respectively.
This amounts to around 80 additionally fixed warnings (across both warning types) per month.
We also observed that Return Not Nullable warnings are usually fixed around twice as fast if an auto-fix is displayed.

Overall, we conclude from our experience of deploying Getafix that automatic repair can be beneficial in practice.
To the best of our knowledge, this is the first industrially deployment of an automated repair tool that automatically learns fix patterns.
Two take-aways for what has made Getafix useful at \doubleblind{CompanyX}{Facebook} are
(i) to integrate auto-fixes into existing development tools, e.g., by showing them in the code-review tool, and
(ii) to predict fixes fast enough to not slow down the development process, e.g., by ensuring that suggestions do not add much to the time a developer waits anyway for the static analysis reports.

\section{Related Work}\label{sec:related}

\subsection{Automated Program Repair}

Getafix has goals similar to those of existing automated program repair techniques~\cite{cacm2019-program-repair}, but fills a previously unoccupied spot in the design space.
In contrast to generate-and-validate approaches, such as GenProg~\cite{Goues2012GenProgAG}, we focus on learning patterns from past fixes for specific kinds of warnings.
Specifically, Getafix does not attempt to find generic solutions from any sort of ingredient space, or by generically mutating the code.
While Genesis~\cite{Long2017Automatic} follows a similar approach of learning transformations between \before and \after ASTs, its learned templates contain generators, which increase the size of the search space.
%
SketchFix~\cite{Hua2018TowardsPP} also aims at validating as few fix candidates as possible, but relies on runtime information from repeated test executions.
In contrast, Getafix does not use any tests.
\citet{Kim2013Automatic} automatically group human-written fixes by similarity. However, inspecting these groups and creating fix patterns from them remains a manual step, whereas Getafix learns fix patterns fully automatically.
History-driven program repair~\cite{Le2016HistoryDP} has a diffing and mining pipeline similar Getafix, but uses it to infer abstract edit steps (like "insert statement"), which then help rank fix candidates produced by a generate-and-validate repair technique.
%
%
Prophet learns a generic model of how natural a fix is and uses it for ranking fix candidates~\cite{Long2016Automatic}.
Their approach could be plugged into Getafix, as an alternative or addition to our ranking algorithm.
Section~\ref{sec:comparison ranking} discusses their ranking in more detail.
CapGen~\cite{Wen2018ContextAwarePG} uses AST context to select mutation operators and fixing ingredients.
Getafix implicitly uses a similar prioritization using its clustering and ranking strategy.

\citet{Soto2016ADeeperLook} observe that fix patterns differ across programming languages and analyze the structure of typical patterns in Java.
Martinez et al.~\cite{Martinez2012Mining,Martinez2015MiningSW,Martinez2018Coming} also performed extensive AST analysis on Java repositories to statistically analyze code change patterns, which can guide program repair.
\citet{Soto2018UsingProbM} propose a probabilistic model-based approach for Java, which produces higher quality fixes more likely. 
As Getafix works on the level of ASTs and learns from human fixes, it can learn language-specific fix patterns automatically.
NPEfix~\cite{cornu2015npefix} addresses NullPointerExceptions by selecting at runtime from a set of manually defined strategies, such as skipping the statement or replacing the null values with a fresh object.
In contrast, Getafix addresses arbitrary bug categories and applies fixes statically.


\subsection{Mining of Edit Patterns}

Refazer~\cite{Rolim2017LearningSPT} uses PROSE~\cite{Polozov2015FlashMeta} and a DSL of program transformations to learn repetitive edits.
Revisar~\cite{Rolim2018LearningQF} influenced Getafix by also using anti-unification~\cite{Kutsia2014AntiUnification} to derive edit patterns from concrete edits, but based on greedy clustering and without extracting surrounding context.
Another difference is that Revisar learns from arbitrary code changes, not fixes of specific bug categories, requiring manual selection of interesting patterns from hundreds of candidates.
See Sections~\ref{sec:comparison greedy} and~\ref{sec:eval ablation} for a detailed comparison.
\citet{Brown2017a} mine recurring bug-introducing changes as the reverse operations of commits.
Their goal is to infer mutation operations, i.e., code changes that inject artificial bugs into arbitrary code, whereas Getafix aims at fixing bugs.

\subsection{Machine Learning on Code}

Several approaches delegate parts of or the entire fix generation task to a neural network.
SSC combines a set of manually written generators of fix candidates with learned models that decide which candidate to apply~\cite{Devlin2017}.
Sarfgen uses embeddings of code to formulate repair as a search for a similar reference solution~\cite{Wang2018}.
Instead, Getafix aims at suggesting fixes for newly developed software that has no reference solution.
DeepFix learns an end-to-end model for predicting fixes of compilation errors~\cite{Gupta2017}.
Combing end-to-end learning with our idea of focusing on fixes for particular bug categories may be interesting future work.

Getafix relates to a stream of work on machine learning applied to code~\cite{Allamanis2018}.
\citet{Pradel2018DeepBugs} demonstrate the power of identifier names for reasoning about buggy code.
A similar technique could benefit our learning and ranking phase, e.g., by using word embeddings of identifiers and literals to guide which variables to substitute holes with, or by mimicking human intuition about return values to pick in case of an early return.
Work on learning to represent code edits may also provide a starting point for learning to suggest fixes~\cite{Yin2018}.

\section{Conclusion}

Fixing incorrect code is a long-standing problem that consumes significant developer time.
Fortunately, many fixes, in particular bugs identified by static analyzers, are instances of recurring patterns.
This paper presents Getafix, the first industrially deployed automated repair tool that fully automatically learns fix patterns from past, human-written commits.
The approach is based on a novel hierarchical, agglomerative clustering algorithm, which summarizes a given set of fix commits into a hierarchy of fix patterns.
Based on this hierarchy and a ranking technique that decides which fix pattern is most appropriate for a new occurrence of a bug category, Getafix proposes a ranked list of fixes to the developer.
An evaluation with 1,268 real-world bug fixes and our experience of deploying Getafix within \doubleblind{a major software company}{Facebook} show that the approach accurately predicts human-like fixes for various bugs, reducing the time developers have to spend on fixing recurring kinds of bugs.

We believe that the potential of the Getafix approach is broader than fixing only static analysis bugs.  As long
as the bug category and bug location is known, and a training set of fixes is available,
Getafix can offer fixes learned from the training set.
Case in point,
Getafix suggests fixes --- via SapFix\doubleblind{}{\footnote{\urlx{https://code.fb.com/developer-tools/finding-and-fixing-software-bugs-automatically-with-sapfix-and-sapienz/}}}~\cite{Marginean2019Sapfix} --- for
null pointer exceptions detected by Sapienz\doubleblind{}{\footnote{\urlx{https://code.fb.com/developer-tools/sapienz-intelligent-automated-software-testing-at-scale/}}}, an automated testing system for mobile apps.

\begin{acks}
  We thank fellow Facebook colleagues for their contributions and help (chronologically):
  Eric Lippert implemented a custom version of the GumTree tree differencing algorithm and investigated an edit script graph approach to learning code change patterns.
  Jeremy Dubreil from the Infer team worked with us to integrate Infer fix suggestions into the existing Infer bug reporting workflow.
  Alexandru Marginean from the Sapienz team integrated Getafix into the SapFix pipeline as one of its strategies to create fix candidates.
  Austin Luo applied Getafix to Hack (programming language), working on learning lint patterns from past change suggestions made during code review.
  Waruna Yapa applied Getafix to Objective-C, working on learning fixes for further classes of Infer warnings (for example "Bad Pointer Comparison").
  Vijayaraghavan Murali created and trained a classifier that provides an alternative approach for ranking between fix patterns.
\end{acks}

\bibliography{references}


\end{document}